\documentclass[usenatbib]{mn2e}

\usepackage{apjfonts,amsfonts,verbatim}
\usepackage{bm}
\usepackage{mathtools}
\usepackage{xcolor}
\usepackage{natbib}
\usepackage{lscape}
\usepackage{afterpage}  
\usepackage{booktabs}
\usepackage{tabularx}
\usepackage{adjustbox}

\usepackage{threeparttable}

\usepackage[T1]{fontenc}

\usepackage{graphicx}	
\usepackage{amsmath}	
\usepackage{amssymb}	

\newcommand{\lya}{Ly$\alpha$}
\newcommand{\ha}{H$\alpha$}
\newcommand{\hb}{H$\beta$}
\newcommand{\oiii}{[\rm O\,{\textsc {iii}}]}

\newcommand{\HI}{\rm H\,{\textsc {i}}}

\newcommand{\kms}{\,\ifmmode{\mathrm{km}\,\mathrm{s}^{-1}}\else km\,s${}^{-1}$\fi}

\title[Deciphering the \lya\ Emission Line]{Deciphering the Lyman-$\alpha$ Emission Line: Towards the Understanding of Galactic Properties Extracted from \lya\ Spectra via Radiative Transfer Modeling}

\author[Li et al.]{Zhihui Li$^{1}$\thanks{E-mail: zhihui@caltech.edu}
and Max Gronke$^{2,3}$\thanks{Hubble fellow}
\\
$^{1}$Cahill Center for Astrophysics, California Institute of Technology, MC 249-17, 1200 East California Boulevard, Pasadena, CA 91125, USA\\
$^{2}$Max-Planck Institute for Astrophysics, Karl-Schwarzschild-Str. 1, D-85741 Garching, Germany\\
$^{3}$Department of Physics \& Astronomy, Johns Hopkins University, Baltimore, MD 21218, USA
}

\date{}

\begin{document}
\label{firstpage}
\pagerange{\pageref{firstpage}--\pageref{lastpage}}
\maketitle

\begin{abstract}
Existing ubiquitously in the Universe with the highest luminosity, the Lyman-$\alpha$ (\lya) emission line encodes abundant physical information about the gaseous medium it interacts with in its diverse morphology. Nevertheless, the resonant nature of the \lya\ line complicates the radiative transfer (RT) modeling of the line profile, making the extraction of physical properties of the surrounding gaseous medium notoriously difficult. In this paper, we revisit the problem of deciphering the \lya\ emission line with RT modeling. We reveal intrinsic parameter degeneracies in the widely-used shell model in the optically thick regime for both static and outflowing cases, which suggest the limitations of the model. We have also explored the connection between the more physically realistic multiphase, clumpy model and the shell model. We find that the parameters of a ``very clumpy'' slab model and the shell model have the following correspondences: (1) the \emph{total} column density of the clumpy slab model is equal to the \HI\ column density of the shell model; (2) the effective temperature of the clumpy slab model, which incorporates the clump velocity dispersion, is equal to the effective temperature of the shell model; (3) the \emph{average} radial clump outflow velocity is equal to the shell expansion velocity; (4) large intrinsic line widths are required in the shell model to reproduce the wings of the clumpy slab models; (5) adding another phase of hot inter-clump medium will increase peak separation, and the fitted shell expansion velocity lies between the outflow velocities of two phases of gas. Our results provide a viable solution to the major discrepancies associated with \lya\ fitting reported in previous literature, and emphasize the importance of utilizing information from additional observations to break the intrinsic degeneracies as well as interpreting the model parameters in a more physically realistic context.

\end{abstract}


\section{Introduction}
Owing to its luminous nature, Lyman-$\alpha$ (\lya) is one of the best emission lines to explore the high-redshift universe, including identifying and studying the formation of distant galaxies as well as probing the reionization era (see a recent review by \citealp{Ouchi20}). Despite all its advantages, the \lya\ line is a resonant transition with a large cross-section, making its radiative transfer (RT) process notoriously difficult to model. Initially, the \lya\ radiative transfer problem was studied analytically for several simple cases, e.g. static plane-parallel slabs \citep{Harrington73, Neufeld90}, a two-phase ISM \citep{Neufeld91}, static uniform spherical shells \citep{Dijkstra06a}, and a uniform neutral IGM with pure Hubble expansion \citep{Loeb99}. Later on, more and more studies started to employ numerical (mostly Monte Carlo) methods in more sophisticated configurations, e.g. flattened, axially symmetric, rotating clouds \citep{Zheng02}, expanding/contracting spherical shells \citep{Zheng02, Ahn03, Ahn04, Dijkstra06a, Dijkstra06b, Verhamme06, Gronke15}, (moving) multiphase, clumpy medium \citep{Richling03, Hansen06, Dijkstra12, Laursen13, Duval14, Gronke16_model}, and anisotropic gas distributions \citep{Behrens14,Zheng14}, as well as in the context of cosmological simulations \citep{Cantalupo05, Tasitsiomi06, Laursen07, Verhamme12}.

With the significant advancements on the theoretical side, many attempts have been made to bridge the gap between the simulations and observations, one of which is to match the \lya\ spectra derived from the RT models with the observed \lya\ profiles. The most widely used RT model for this endeavor is the `shell model', i.e. a spherical, expanding/contracting \HI\ shell. Thus far, the shell model has managed to reproduce a wide variety of \lya\ profiles, including typical single and double-peaked profiles from Lyman break galaxies (LBGs), \lya\ emitters (LAEs), damped \lya\ systems (DLAs) and Green Pea galaxies (e.g. \citealp{Verhamme08, DZ10, Vanzella10, Krogager13, Hashimoto15, Yang16, Yang17a}), along with the P-Cygni profiles and damped absorption features in nearby starburst galaxies \citep[e.g.][]{Atek09, Martin15}. Nevertheless, a number of discrepancies between the fitted parameters of the shell model and observational constraints have been observed \citep[e.g.][]{Kulas12, Hashimoto15, Yang16, Yang17a}. Most recently, \citet{Orlitova18} reported three major discrepancies emerged from shell modeling of the observed \lya\ profiles of twelve Green Pea galaxies, namely: (1) the required intrinsic \lya\ line widths are on average three times broader than the observed Balmer lines; (2) the inferred outflow velocities of the shell ($\lesssim$\,150\,km\,s$^{-1}$) are significantly lower than the characteristic outflow velocities ($\sim$\,300\,km\,s$^{-1}$) indicated by the observed ultraviolet (UV) absorption lines of low-ionization-state elements; (3) the best-fit systemic redshifts are larger (by 10 – 250\,km\,s$^{-1}$) than those derived from optical emission lines. Such inconsistencies suggest the limitations of the shell model and necessitate the development of more realistic RT models.

In addition, it is unclear whether the derived values of the shell model can be directly used to infer other physical properties of the \lya-emitting object. For example, \citet{Verhamme15} proposed that low \HI\ column densities ($\lesssim 10^{18}\,{\rm cm}^{-2}$) inferred from observed \lya\ profiles should indicate Lyman-continuum (LyC) leakage. However, it has not been verified quantitatively that a tight correlation 
does exist between the \HI\ column density inferred from \lya\ and the LyC escape fraction as expected theoretically (see Eq. (4) in \citealt{Verhamme17}). The situation is even more complicated when more physics (e.g. turbulence) is considered, e.g. \citet{Kakiichi21} find that a high average \HI\ column density still allows high LyC leakage, as LyC photons can escape through narrow photoionized channels with a large fraction of hydrogen remaining neutral (see their Section 4.2; see also \citealt{Kimm19}).

The shell model is known as being unrealistically monolithic as it consists of only one phase of \HI\ at $\sim 10^4\,$K (the `cool' phase). Alternatively, \lya\ radiative transfer has been studied in multiphase, clumpy models (e.g. \citealt{Neufeld91, Hansen06, Dijkstra12, Laursen13, Duval14, Gronke16_model}), as numerous observations have revealed the multiphase nature of the interstellar/circumgalactic/intergalactic medium (ISM/CGM/IGM, respectively; see reviews by \citealt{Cox05, Tumlinson17, McQuinn16}). This multiphase, clumpy model consists of two different phases of gas: cool clumps of \HI\ ($\sim 10^4\,$K) embedded in a hot, highly-ionized medium ($\sim 10^6\,$K). Using the framework in \citet{Gronke16_model}, \citet{Li21} and \citet{Li21b} successfully reproduced the spatially-resolved \lya\ profiles in \lya\ blobs 1 and 2 with the multiphase, clumpy model. These results have not only demonstrated the feasibility of the multiphase, clumpy model, but also motivated us to gain a deeper understanding of the physical meaning of the derived model parameters.

The primary goal of this work is to figure out the links between the parameters of the relatively newly-developed, more physically realistic multiphase, clumpy model and the commonly-adopted shell model, as well as what physical information can be extracted from observed \lya\ spectra. The shell model only has four most important parameters\footnote{Here we assume that the shell model is dust-free, as the dust content is usually poorly constrained by the observed \lya\ spectra \citep{Gronke15}.}: the shell expansion velocity ($v_{\rm exp}$), the shell \HI\ column density ($N_{\rm HI,\,shell}$), the shell effective temperature ($T_{\rm shell}$) or the Doppler parameter ($b$), and the intrinsic \lya\ line width ($\sigma_{\rm i}$) \citep{Verhamme06,Gronke15}. Similarly, the multiphase, clumpy model also has four most crucial  parameters: (1) the residual \HI\ number density in the inter-clump medium (ICM, $n_{\rm HI,\,{\rm ICM}}$); (2) the cloud covering factor ($f_{\rm {\rm cl}}$), which is the mean number of clumps per line-of-sight; (3) the velocity dispersion of the clumps ($\sigma_{\rm {\rm cl}}$); (4) the radial outflow velocity of the clumps ($v_{\rm cl}$). For both models, an additional post-processed parameter, $\Delta v$, is used to determine the systemic redshift of the \lya\ emitting source. The shell model parameters capture different properties of the \lya\ spectra: $v_{\rm exp}$ determines the red-to-blue peak flux ratio, and sets the position of the absorption trough between two peaks (as $-v_{\rm exp}$ corresponds to the largest optical depth); $N_{\rm HI,\,shell}$ dictates the amount of peak separation and the depth of the absorption trough; $T_{\rm shell}$ or $b$ describes the internal kinematics of the shell (including thermal and turbulent velocities) and controls the width of the \lya\ profile, but is usually poorly constrained by the data \citep{Gronke15}; $\sigma_{\rm i}$ (if large enough) sets the extent of the wings of the spectrum. The multiphase, clumpy model parameters capture similar spectral properties but in different ways: $v_{\rm cl}$ determines the red-to-blue peak flux ratio\footnote{In the multiphase, clumpy model, the absorption trough is not necessarily set by $-v_{\rm cl}$ unless the total column density of the clumps is high enough to be optically thick (i.e. the flux density at line center is close to zero) and the clumps and ICM are co-outflowing at the same velocity (see \S\ref{sec:multiphase_slab}).}; $n_{\rm HI,\,{\rm ICM}}$ and $f_{\rm {\rm cl}}$ together dictate the amount of peak separation and the depth of the absorption trough, as both of them contribute to the total \HI\ column density; $\sigma_{\rm cl}$ sets the width of the spectrum.

\citet{Li21b} have observed significant correlations between pairs of model parameters (namely $\sigma_{\rm i} - \sigma_{\rm cl}$ and $v_{\rm exp} - v_{\rm cl}$) derived by fitting fifteen observed \lya\ spectra. These results are enlightening yet not rigorous and may suffer from parameter degeneracy due to their empirical nature. Motivated by the fact that the multiphase, clumpy model may converge to the shell model in the limit of very high $f_{\rm {\rm cl}}$ \citep{Gronke17b}, in this work we attempt to find quantitative correlations between the parameters of two models, with the aim of better understanding the physical meaning of model parameters and their relation to \lya\ spectral properties.

The structure of this paper is as follows. In \S\ref{sec:method}, we describe the methodology of this work. In \S\ref{sec:degeneracy}, we present the intrinsic parameter degeneracies of the shell model. In \S\ref{sec:results}, we explore the connection between the shell model and the multiphase, clumpy model. In \S\ref{sec:discussion}, we discuss on how to interpret the physical parameters extracted from \lya\ spectra. In \S\ref{sec:conclusion}, we summarize and conclude. The physical units used throughout this paper are km\,s$^{-1}$ for velocity, cm$^{-2}$ for column density, and K for temperature, unless otherwise specified.

\section{Methodology}\label{sec:method}
In this work, we extract physical parameters from \lya\ spectra by fitting them with a grid of shell models. The fitted \lya\ spectra can be one of the following: (1) a shell model spectrum; (2) a (multiphase) clumpy model spectrum; (3) an observed \lya\ spectrum. The grid of shell models that we use was previously described in \citet{Gronke15}. This shell model grid consists of 12960 discrete RT models, with [$v_{\rm exp}$, ${\rm log}\,N_{\rm HI,\,shell}$, ${\rm log}\,T_{\rm shell}$] varying between [0, 490] km\,s$^{-1}$, [16.0, 21.8] cm$^{-2}$ and [3, 5.8] K, respectively. Each shell model is calculated via Monte-Carlo RT using $20000$ \lya\ photon packages generated from an a priori Gaussian intrinsic spectrum \emph{N}(0, $\sigma^2$), where $\sigma$ = 800\,km\,s$^{-1}$. The intrinsic \lya\ line width, $\sigma_{\rm i} \in [1, 800]\,\rm km\,s^{-1}$, is accounted for in the form of a weighting function in post-processing. We do not consider the effect of dust in this work as it is usually a poorly constrained parameter \citep{Gronke15}.

To properly explore the possibly multimodal posterior of the shell model parameters, we use a python package \texttt{dynesty} \citep{Skilling04, Skilling06, Speagle20} that implements the nested sampling algorithm for our fitting pipeline. The model spectrum of each sampled point in the parameter space is calculated via linear flux interpolation on the model grid rather than running the computationally expensive RT ``on the fly''. When fitting the \lya\ spectra, we manually add a constant 1-$\sigma$ uncertainty of about 10\% of the maximum flux density to the normalized (mock) data to reflect the typical observational uncertainties.

\section{Results I: Intrinsic Parameter Degeneracies of the Shell Model}\label{sec:degeneracy}

In this section, we show the existence of intrinsic parameter degeneracies in the shell model revealed by fitting, in preparation for our subsequent discussion. We consider the two following cases: static shell and outflowing shell, respectively.

\subsection{Static Shells: Degeneracy between $N_{\rm HI,\,shell}$ and $T_{\rm shell}$ }\label{sec:static_deg}

Here we show that for static shells in the optically thick regime, models with the same $N_{\rm HI,\,shell}\,T_{\rm shell}^{0.5}$ exhibit identical \lya\ spectra. Theoretically, the angular-averaged \lya\ spectral intensity $J(x)$ emerging from a static, uniform \HI\ sphere is given analytically as \citep{Adams72,Harrington73,Neufeld90,Dijkstra06a}:

\begin{equation}\label{eq:neufeld}
J(x)=\frac{\sqrt{\pi}}{\sqrt{24}a\tau_0}\Bigg{(}\frac{x^2}{1+{\rm
cosh}\Big{[}\sqrt{\frac{2\pi^3}{27}}\frac{|x^3|}{a\tau_0}\Big{]}}\Bigg{)}
\end{equation} 
where $x\equiv (\nu-\nu_0)/\Delta \nu_D$ is the unitless frequency, and the Doppler parameter $\Delta \nu_D = v_{\rm th}\nu_0/c = \sqrt{2k_{\rm B}T/m_{\rm H}}\nu_0/c$, with $T$ being the \HI\ gas temperature. Here $\nu$ is the \lya\ photon frequency and $\nu_0=2.47 \times 10^{15}$ Hz is the \lya\ central frequency. Moreover, $a = \Delta\nu_{\rm L} / 2\Delta \nu_{\rm D} \propto T^{-0.5}$ is the Voigt parameter, where $\Delta\nu_{\rm L}$ is the natural line broadening; $\tau_{0}$ is the \HI\ optical depth at the line center and $\tau_{0} \propto N_{\rm HI}\,T^{-0.5}$. The complete expressions for $a$ and $\tau_{0}$ can be found, e.g. in \citet{Dijkstra06a}.

One can then switch from the frequency space to the velocity space by converting $J(x)$ to $J(v)$ via $(\nu-\nu_0)/\nu_0 = x v_{\rm th}/c$. Then it is evident that with proper normalization, $J(v)$ would be identical for different combinations of $(N_{\rm HI}, T)$ that give the same $a \tau_{0} v_{\rm th}^{3}$, which is $\propto N_{\rm HI}\,T^{0.5}$. Alternatively, one can derive this $N_{\rm HI}\,T^{0.5}$ degeneracy by estimating the most likely escape frequency of \lya\ photons (see e.g. Eq. (5) in \citealt{Gronke17b}, which originally comes from \citealt{Adams72}).

\begin{figure}
\centering
\includegraphics[width=0.47\textwidth]{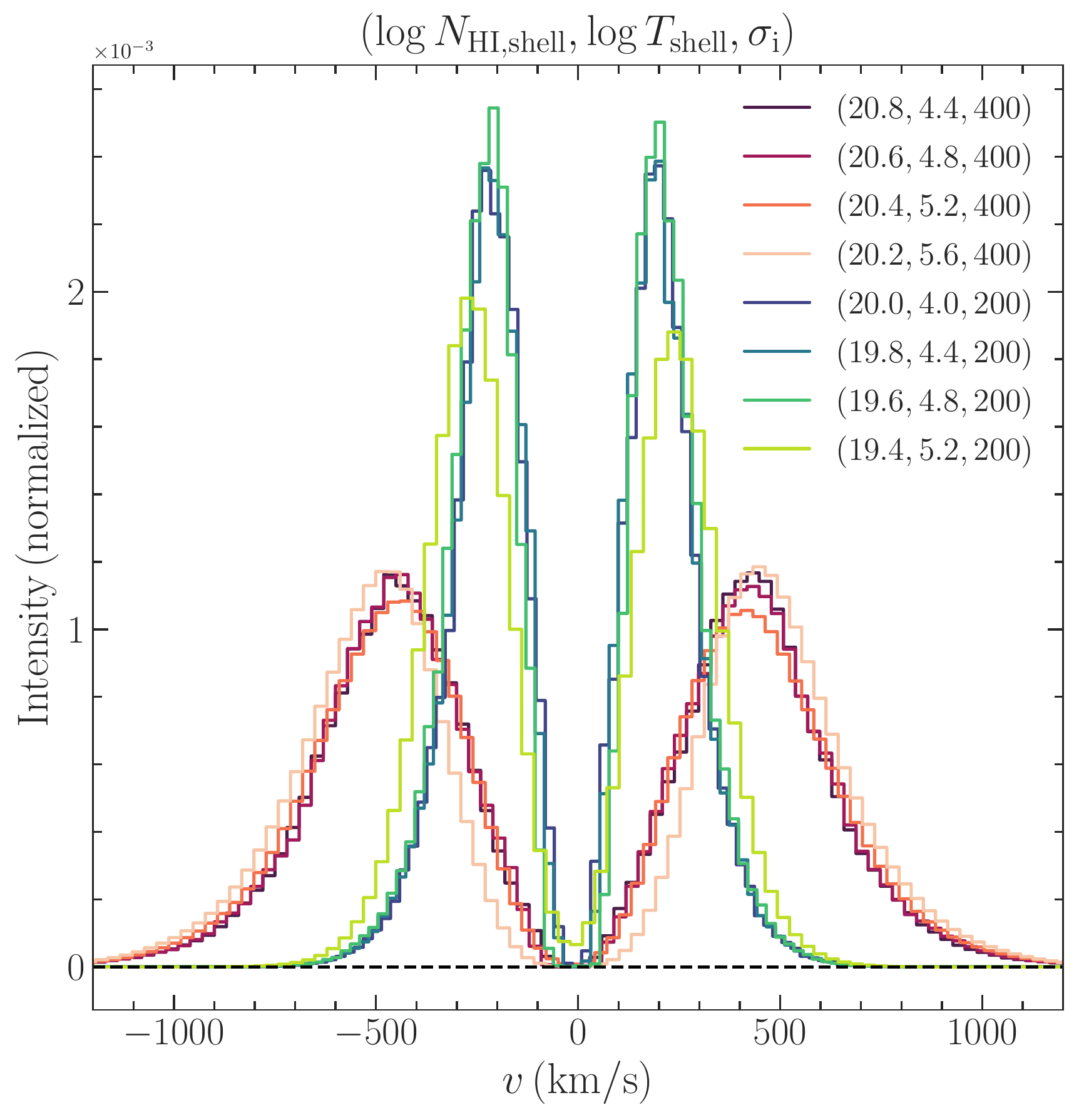}
    \caption{\textbf{Examples of degenerate static shell models with the same $N_{\rm HI}\,T^{0.5}$.} Different colored curves represent two sets of shell models ($\sigma_{\rm i} =$ 200 and 400 $\rm km\,s^{-1}$, respectively) with the same $N_{\rm HI}\,T^{0.5}$. It can be seen that the normalized intensity distributions of each set of models are nearly identical. Note that this degeneracy only exists in the optically thick regime ($a \tau_{0} \gtrsim 10^3$); at $a \tau_{0} = 2.8 \times 10^{-13} (N_{\rm HI}/T) \simeq 100$ the models start to deviate from the other degenerate models (the light lime and light pink curves).
    \label{fig:degeneracy}}
\end{figure}

\begin{figure}
\centering
\includegraphics[width=0.47\textwidth]{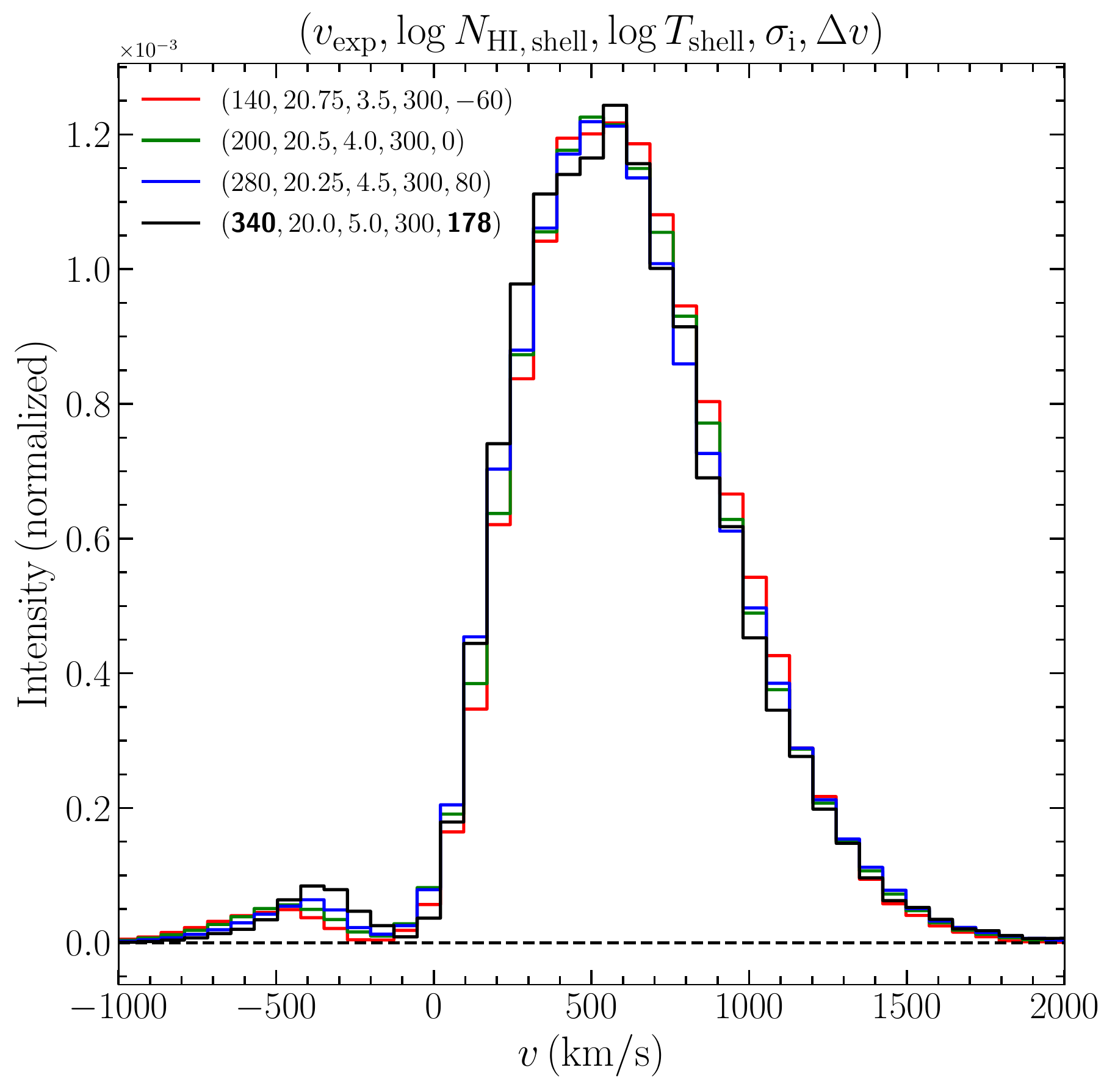}
\includegraphics[width=0.47\textwidth]{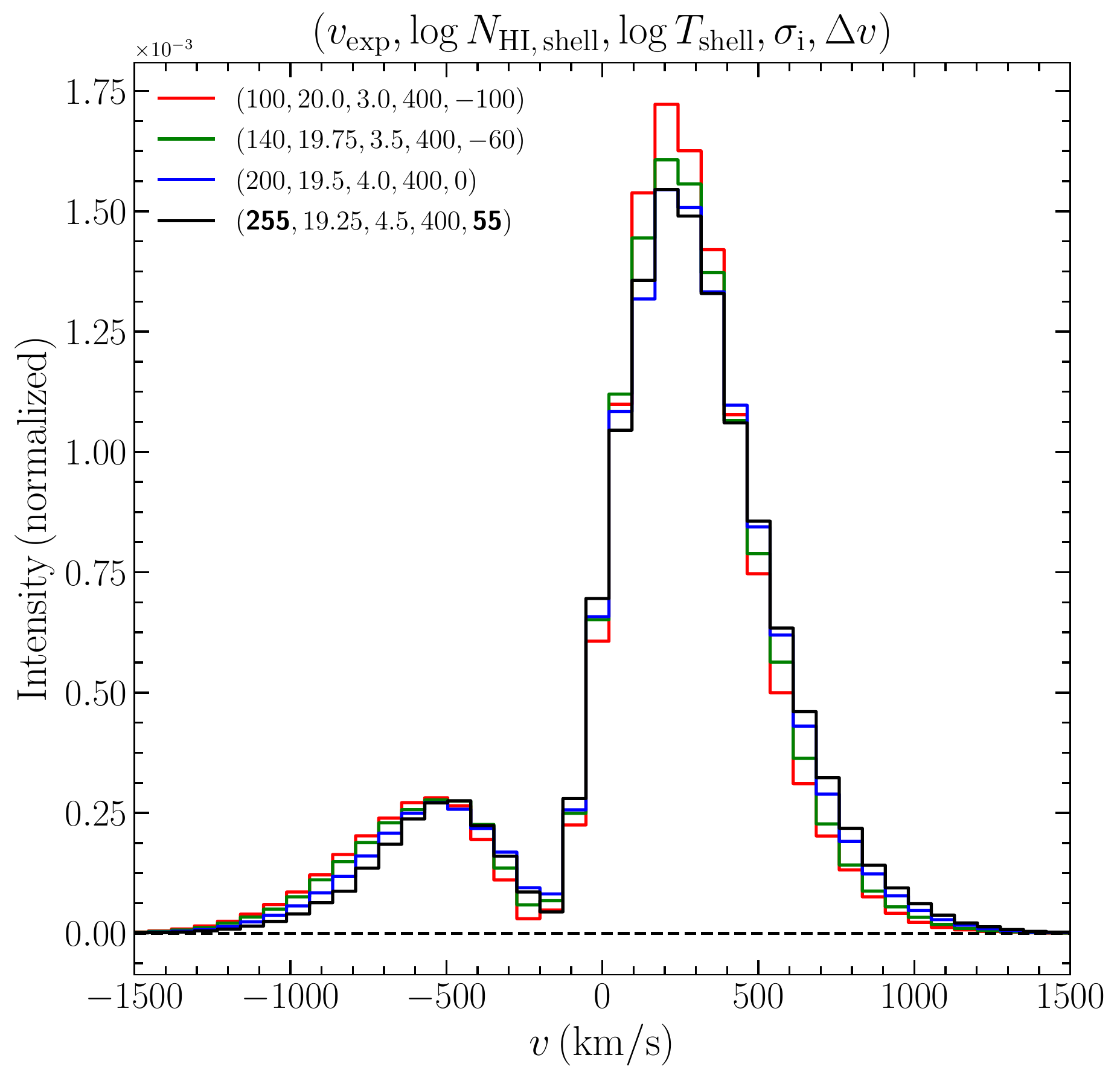}
    \caption{\textbf{Examples of degenerate outflowing shell models.} Different colored curves represent two sets of outflowing shell models ($\sigma_{\rm i} =$ 300 and 400 $\rm km\,s^{-1}$, respectively), each consisting of a series of models with increasing $T_{\rm shell}$, with a step size of 0.5 dex. Accordingly, $N_{\rm HI,\,shell}$ decreases by 0.25 dex and $v_{\rm exp}$ increases by a factor of $\sqrt{2}$. Each set of spectra appear essentially identical to each other, except for the one with the largest $T_{\rm shell}$ (the black curve), which is the only model in the series that does not satisfy $a \tau_{0} \gtrsim 10^3$. We fit this model with our shell model grid by fixing $N_{\rm HI,\,shell}$, $T_{\rm shell}$ and $\sigma_{\rm i}$ at the expected values and leaving $v_{\rm exp}$ and $\Delta v$ free. A decent fit is achieved, albeit with the best-fit $v_{\rm exp}$ values (shown in bold) slightly lower than expected.
    \label{fig:degeneracy_v}}
\end{figure}

We show this degeneracy in Figure \ref{fig:degeneracy} for two sets of static shell models ($\sigma_{\rm i} =$ 200 and 400 $\rm km\,s^{-1}$, respectively) with the same $N_{\rm HI}\,T^{0.5}$. It can be seen that the normalized intensity distributions of each set of models are nearly identical (modulo numerical noise). Note that this degeneracy only exists in the optically thick regime (i.e. $a \tau_{0} \gtrsim 10^3$) where Eq. (\ref{eq:neufeld}) holds \citep{Harrington73,Neufeld90}. As shown in Figure \ref{fig:degeneracy}, models with the same $N_{\rm HI}\,T^{0.5}$ but $a \tau_{0} = 2.8 \times 10^{-13} (N_{\rm HI}/T) \simeq 100$ start to deviate from the other degenerate models, as Eq. (\ref{eq:neufeld}) is no longer applicable.

\subsection{Outflowing Shells: Degeneracy among ($v_{\rm exp}, N_{\rm HI,\,shell}, T_{\rm shell}, \Delta v$)}\label{sec:outflow_deg}

If the shell is outflowing, the $N_{\rm HI}\,T^{0.5}$ degeneracy starts to be broken -- in fact, the models with higher $N_{\rm HI}$ will have fewer flux in the blue peak, as it is more difficult for the blue photons to escape from the shell. However, such a larger level of asymmetry can be compensated by a lower shell expansion velocity. Heuristically, we find that if we allow the \lya\ spectra to shift along the velocity axis (i.e. the systemic velocity of \lya\ source is not necessarily at zero; this is often the case for fitting real observed \lya\ spectra, where the systemic redshift of the \lya\ source has considerable uncertainties), two shell models with $(v_{\rm exp}, {\rm log}\,N_{\rm HI,\,shell}, {\rm log}\,T_{\rm shell})$ and $\sim (2v_{\rm exp}, {\rm log}\,N_{\rm HI,\,shell} - 0.5\,{\rm dex}, {\rm log}\,T_{\rm shell} + 1\,{\rm dex}, \Delta v)$ are degenerate with each other, where $\Delta v$ is the difference in systemic velocity of the two \lya\ sources. We have not been able to analytically derive such a quadruple parameter degeneracy rigorously, but we verify its existence numerically in this section. 

We show this degeneracy with two sets of examples in Figure \ref{fig:degeneracy_v}. Each set contains a series of models with increasing $T_{\rm shell}$, with a step size of 0.5 dex. Accordingly, $N_{\rm HI,\,shell}$ decreases by 0.25 dex and $v_{\rm exp}$ increases by a factor of $\sqrt{2}$. As can be seen in Figure \ref{fig:degeneracy_v}, each set of spectra appear essentially identical to each other, except for the one with the largest $T_{\rm shell}$ (the black curve), which is the only model in the series that does not satisfy $a \tau_{0} \gtrsim 10^3$. We fit this model with our shell model grid by fixing $N_{\rm HI,\,shell}$, $T_{\rm shell}$ and $\sigma_{\rm i}$ at the expected values and leaving $v_{\rm exp}$ and $\Delta v$ free. It turns out that a decent fit can be achieved, with the best-fit $v_{\rm exp}$ values (shown in bold) slightly lower than expected. In other words, this quadruple degeneracy is broken quantitatively but still holds qualitatively.

Such a quadruple degeneracy reminds us of the limitation of shell models in fitting observed \lya\ spectra, as $v_{\rm exp}$, $N_{\rm HI,\,shell}$, $T_{\rm shell}$ and $\Delta v$ cannot be determined independently by merely fitting. Additional constraints (e.g. a very accurate measurement of the systemic redshift of the \lya\ emitting source) have to be introduced break the parameter degeneracy.

\subsubsection{A Real-World Example: Fitting the \lya\ Spectrum of a Green Pea Galaxy, GP 0911+1831}\label{sec:degeneracy_example}

Here we further show the quadruple degeneracy with a practical example. We fit an observed \lya\ spectrum of a Green Pea galaxy, GP 0911+1831 ($z$ = 0.262236; \citealt{Henry15}) with our shell model grid. The spectrum is obtained from the \lya\ Spectral Database (LASD\footnote{\url{http://lasd.lyman-alpha.com}}; \citealt{LASD}). Following \citet{Orlitova18}, we account for the spectral resolution of the HST Cosmic Origins Spectrograph (COS) by convolving the shell model spectra with a FWHM = 100 $\rm km\,s^{-1}$ Gaussian before comparing them to the observed \lya\ spectrum\footnote{Note that different from \citet{Orlitova18}, we do not consider the effect of dust, as the dust optical depth is usually a poorly constrained parameter and may introduce additional degeneracy \citep{Gronke15}.}. 

\begin{figure}
\centering
\includegraphics[width=0.47\textwidth]{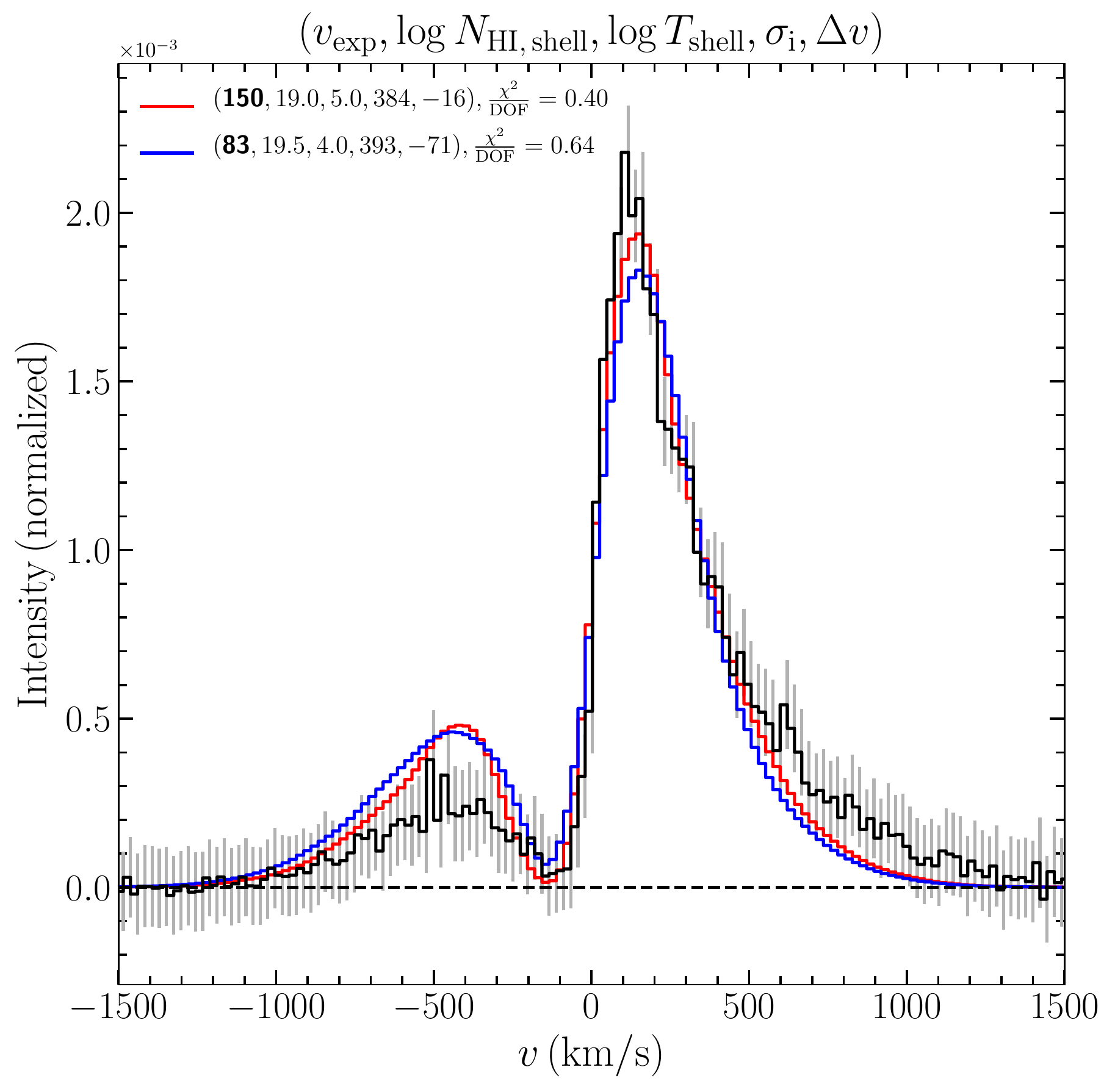}
    \caption{\textbf{Degeneracy shown in fitting the \lya\ spectrum of a Green Pea galaxy, GP 0911+1831.} The observed spectrum is shown in black and two degenerate best-fit shell models are shown in red and blue, respectively. The $\chi^2$ (per degree of freedom, DOF) values of these two best-fit models are very close to each other, but the shell expansion velocity of the high temperature model is about a factor of two higher than the low temperature model (as highlighted in bold), which consequently affects the fitted systemic redshift of the \lya\ source. This result may explain the two major discrepancies reported in the literature (see \S\ref{sec:degeneracy_example} for details).
    \label{fig:degeneracy_example}}
\end{figure}

We present two degenerate best-fit shell models in Figure \ref{fig:degeneracy_example}. These two best-fit models, whose $\chi^2$ per degree of freedom are very close to each other, have the parameter degeneracy as described in \S\ref{sec:outflow_deg} -- the shell expansion velocity of the high temperature model is about a factor of two higher than the low temperature model, which consequently affects the fitted systemic redshift of the \lya\ source. This result may explain the two major discrepancies reported in \citet{Orlitova18}: (1) the inferred shell outflow velocities are significantly lower than the characteristic outflow velocities indicated by the observed UV absorption lines; (2) the best-fit systemic redshifts are larger than those derived from optical emission lines. When fitting observed \lya\ spectra, the best-fit model with a low $v_{\rm exp}$ may happen to provide the best match for the data, but another degenerate model (or a series of degenerate models) with much higher $v_{\rm exp}$ values can actually fit the data similarly well and hence should also be considered as reasonable solutions.

\section{Results II: Connecting the Shell Model to the Multiphase, Clumpy Model}\label{sec:results}
In this section, we attempt to connect the shell model parameters to the multiphase, clumpy model parameters. We generate a series of clumpy models as our ``mock data'' for fitting. We first consider a three-dimensional semi-infinite slab geometry (\S\ref{sec:static} -- \ref{sec:multiphase_slab}) and later we will consider a finite spherical geometry (\S\ref{sec:multiphase_sphere}). This is because for a semi-infinite clumpy slab, it is numerically easier to achieve a very high clump covering factor ($f_{\rm cl} \gtrsim 1000$, i.e. the average number of clumps per line-of-sight is large enough to be in the ``very clumpy'' regime, where the clumpy medium is expected to behave like a homogeneous medium in terms of the emergent \lya\ spectrum, \citealt{Gronke17}), which is prohibitively computationally expensive for a finite clumpy sphere. The clumpy slab models are periodic in the $x$ and $y$ directions with a half-height $B$ of 50 pc\footnote{We emphasize that it is the \HI\ column density that actually matters in the radiative transfer instead of the physical scales of the models.} in the $z$ direction. The clumps within the slab are spherical with radius of $r_{\rm cl}$ = 10$^{-3}$\,pc filled with \HI\ of a column density $N_{\rm HI,\,cl}$. The clump covering factor is directly proportional to the volume filling factor of the clumps $F_{\rm V}$ via $f_{\rm cl} = 3 F_{\rm V} B / 4 r_{\rm cl}$ \citep{Dijkstra12,Gronke17b}.

Each clumpy model is calculated via Monte-Carlo RT using $10000$ \lya\ photon packages assuming a Gaussian intrinsic spectrum \emph{N}(0, $\sigma^2_{\rm i}$), where $\sigma_{\rm i}$ = 12.85\,km\,s$^{-1}$ is the canonical thermal velocity dispersion of $T$ = 10$^4$\,K \HI\ gas in the clumps\footnote{In the clumpy model, $\sigma_{\rm i}$ is fixed to be small and the clump velocity dispersion is responsible for the broadening of the spectrum. $\sigma_{\rm i}$ will not affect \lya\ model spectra as long as it is smaller than the clump velocity dispersion (which is almost always the case, see \S\ref{sec:random}).}. Each model spectrum is normalized to a total flux of one before being fitted with the shell model grid.

\subsection{Clumpy Slab with Static Clumps}\label{sec:static}

\begin{figure}
\centering
\includegraphics[width=0.5\textwidth]{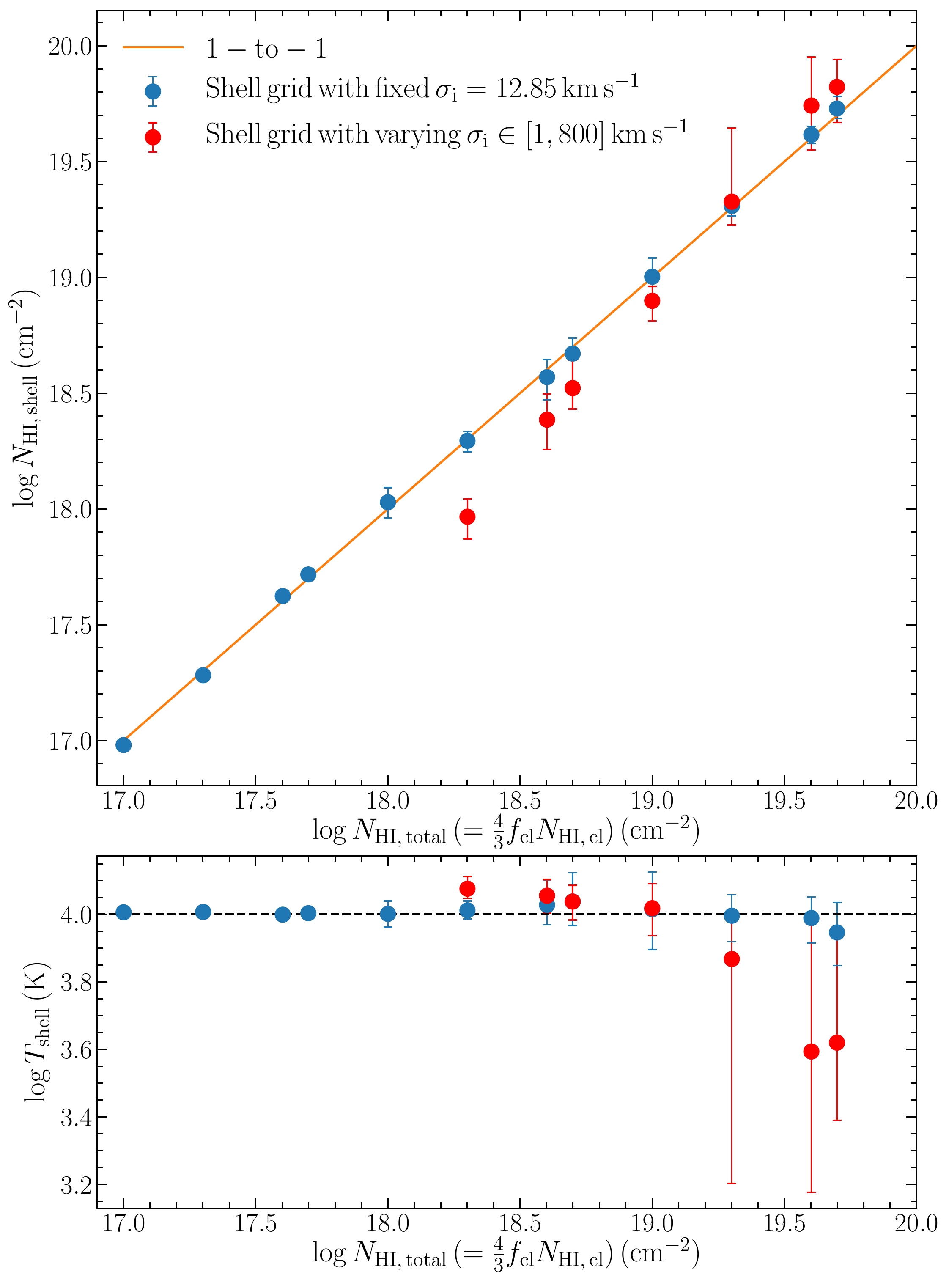}
    \caption{\textbf{Results of fitting static clumpy slab models with static shell models.} The blue and red points represent the parameter values derived from fitting with the customized, small shell model grid (with fixed $\sigma_{\rm i} = 12.85\,\rm km\,s^{-1}$) and the large shell model grid (with varying $\sigma_{\rm i} \in [1, 800]\,\rm km\,s^{-1}$), respectively. \emph{Upper:} The correlation between $N_{\rm HI,\,total}$ and $N_{\rm HI,\,shell}$. A very tight 1-to-1 correlation is present over three orders of magnitude. \emph{Lower:} The distribution of the best-fit shell model temperatures $T_{\rm shell}$. In all cases, $T_{\rm shell}$ values of 10$^{4}$ K (the clump temperature) are obtained within 1-$\sigma$ uncertainties.
    \label{fig:NHI}}
\end{figure}

We start by correlating the \HI\ column density of the shell model with the equivalent average total column density of the single-phase, clumpy slab model, which is given by $N_{\rm HI,\,{\rm total}} = \frac{4}{3}f_{\rm cl}N_{\rm HI,\,cl}$, where the factor $4/3$ comes from the spherical geometry of the clumps \citep{Gronke17b}. We first generate a series of static, single-phase, clumpy slab models by varying $N_{\rm HI,\,cl}$ of the clumps with a very high covering factor $f_{\rm cl}$ (i.e. in the ``very clumpy'' regime) as the mock data. The clumps are fixed to a temperature of $T_{\rm cl} = 10^4\,\rm K$ and do not have any motions (neither random nor outflow velocities). The parameter values that we use are given in the first row of Table \ref{tab:parameter}.

We first attempt to fit the clumpy slab model spectra with the large grid of shell models that we have described in \S\ref{sec:method}. We find that the best-fit shell models are usually noisy and unsatisfactory due to the low number of effective photon packages -- i.e. in order to match the relatively narrow widths of the clumpy slab model spectra (especially the ones with low $N_{\rm HI,\,{\rm total}}$), a weighting function with a small $\sigma_{\rm i}$ ($\lesssim 100\,\rm km\,s^{-1}$, the actual intrinsic \lya\ line width needed) is required, which effectively only includes only a small fraction of modeled photons\footnote{This problem is mitigated when the clumps have a considerable random velocity dispersion, which broadens the spectrum significantly (see \S\ref{sec:random} and the subsequent sections).}. Therefore, we build a customized grid of shell models to fit the clumpy slab model spectra. Such a grid is smaller but similar to the large shell model grid, with two major differences: (1) the shell expansion velocity is fixed to zero; (2) the photon packages are generated from a Gaussian intrinsic spectrum \emph{N}(0, $\sigma_{\rm i}^2$) with $\sigma_{\rm i}$ = 12.85\,km\,s$^{-1}$, i.e. the same as the fitted clumpy slab models. In other words, the intrinsic \lya\ spectrum has a fixed small line width that is also used to generate the mock data. We find that such a customized grid with only two varying parameters [${\rm log}\,N_{\rm HI,\,shell}$, ${\rm log}\,T_{\rm shell}$] can yield better fits (as all the modeled photons contribute to the model spectra) and is significantly faster at fitting the mock data.

\begin{figure}
\centering
\includegraphics[width=0.5\textwidth]{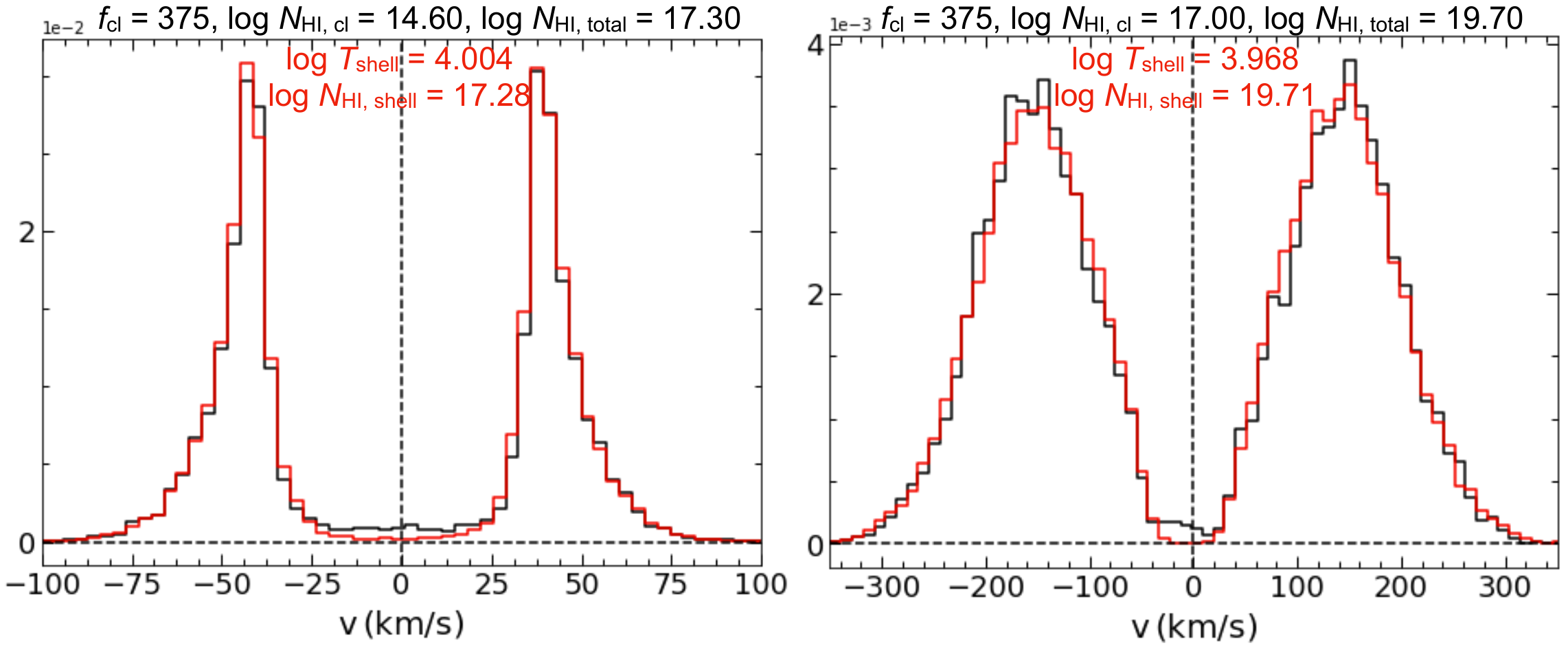}
    \caption{\textbf{Examples of static shell model best-fits (obtained by using the shell model grid with fixed $\sigma_{\rm i}$) to static clumpy slab models.} Two panels represent two different [$f_{\rm cl}, N_{\rm HI,\,cl}$] cases. The black curves represent the static clumpy slab model spectra and the red curves represent the shell model best-fits. Both $T_{\rm shell}$ and $N_{\rm HI,\,shell}$ have been obtained at the expected values.
    \label{fig:static_fits}}
\end{figure}

As shown in Figure \ref{fig:NHI}, there is a tight, 1-to-1 correlation between $N_{\rm HI,\,{\rm total}}$ and $N_{\rm HI,\,{\rm shell}}$ over three orders of magnitude. Moreover, all of the $T_{\rm shell}$ values are consistent with 10$^4$\,K (the clump temperature) within 1-$\sigma$ uncertainties. Therefore, we conclude that the equivalent \HI\ column density of a static, very clumpy slab can be exactly reproduced by a shell model with the same \HI\ column density and the same temperature of the clumps. We show two examples of static shell model best-fits to static clumpy slab models in Figure \ref{fig:static_fits}. 

Despite the shortcomings mentioned above, the large shell model grid is used to fit several static clumpy slab models to verify our results. Several examples\footnote{These examples have $N_{\rm HI,\,{\rm total}}$ high enough to yield $\sigma_{\rm i} \gtrsim 50\,\rm km\,s^{-1}$, below which the fraction of photons included is too low to yield a decent fit.} are shown in Figure \ref{fig:NHI} with red points. We find that $N_{\rm HI,\,{\rm shell}}$ and $T_{\rm shell}$ can still be roughly obtained at their expected values, albeit with small deviations and larger uncertainties. The required intrinsic line widths range from $\sim 50$ to $\sim 100\,\rm km\,s^{-1}$, depending on the width of the mock data. These intrinsic line width values should not have any physical meaning but just ensure that the extent of the wings is proper to yield a good fit. 

\begin{table*}
    \centering
    \caption{Parameter values of the clumpy slab models (the mock data).}
    \label{tab:parameter}
    \setlength{\tabcolsep}{3pt}
    \begin{tabular}{ccccccc}
    \hline\hline
    Model Parameter & $F_{\rm V}$ & $f_{\rm cl}$ & ${\rm log}\,N_{\rm HI,\,{\rm cl}}$ &$\sigma_{\rm cl}$ &$v_{\rm cl}$\\
    \hline
    Definition & Volume filling factor & Clump covering factor & Clump \HI\ column density & Clump random velocity & Clump outflow velocity\\
     (1)  & (2) & (3) & (4) & (5) & (6)\\
    \hline
    Static Clumps & 0.1 & 375 & 14.3 - 17.0 cm$^{-2}$ & 0 & 0\\
    Randomly Moving Clumps & 0.02 - 0.12 & 750 - 4500 & 15.7 - 17.6 cm$^{-2}$ & (50, 100) km\,s$^{-1}$  & 0\\
    Outflowing Clumps & 0.08 - 0.12 & 3000 - 4500 & 15.7 - 17.0 cm$^{-2}$ & (0, 50, 100) km\,s$^{-1}$ & 50 - 400 km\,s$^{-1}$\\
     \hline
    
    \hline\hline
    \end{tabular}
\end{table*}

\subsection{Clumpy Slab with Randomly Moving Clumps}\label{sec:random}
\begin{figure}
\centering
\includegraphics[width=0.47\textwidth]{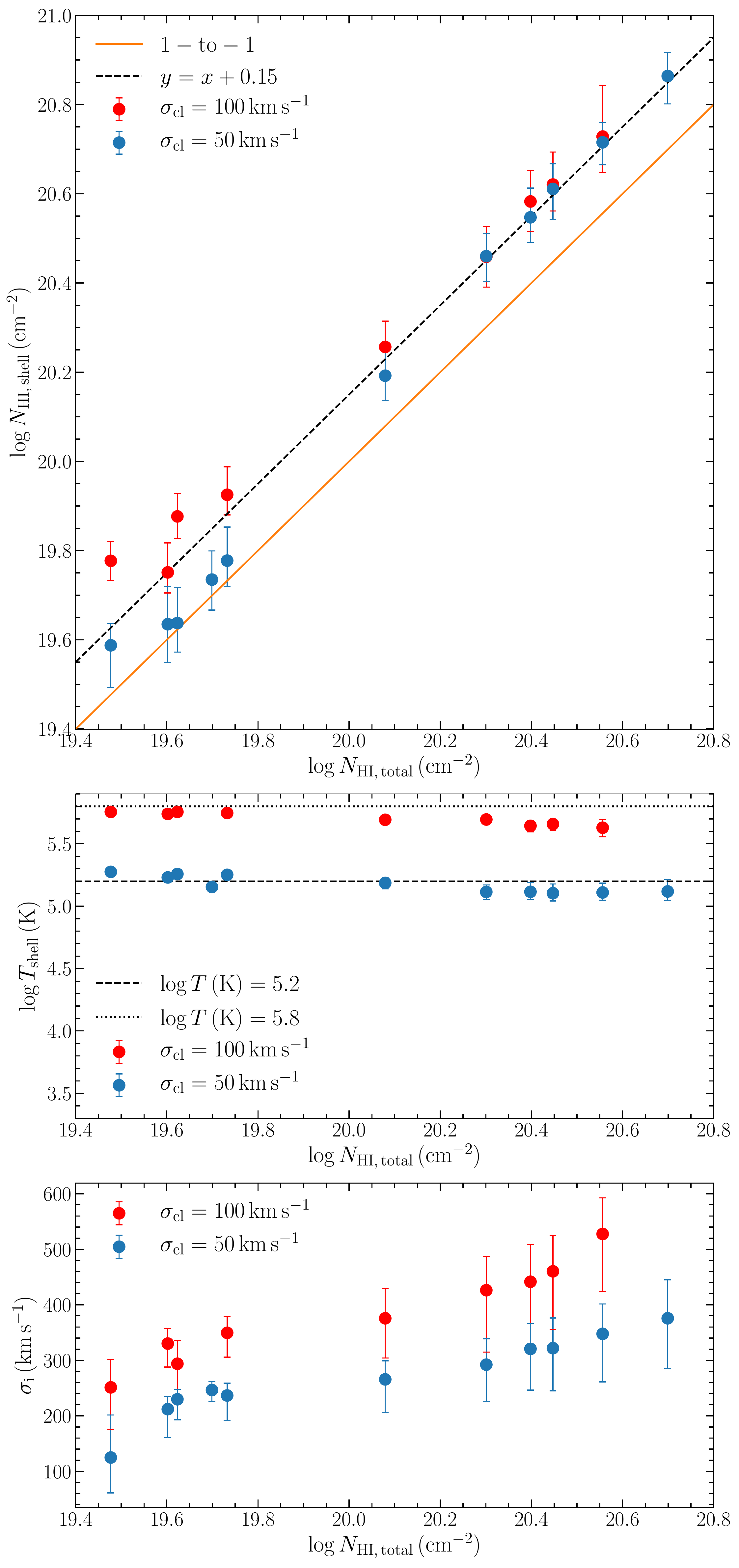}
    \caption{\textbf{Results of fitting clumpy slab models with randomly moving clumps with shell models.} \emph{Upper:} The yielded $N_{\rm HI,\,shell}$ values are around the $N_{\rm HI,\,total}$ values, but a noticeable deviation has emerged. \emph{Middle:}  The yielded shell temperatures ($T_{\rm shell}$) are mostly at the effective temperatures of the clumpy slab model. \emph{Lower:} The distribution of the derived line widths of the intrinsic \lya\ emission ($\sigma_{\rm i}$) of the best-fit shell models. The blue and red points represent the $\sigma_{\rm cl} =$ 50 and 100 $\rm km\,s^{-1}$ models, respectively.
    \label{fig:sigma}}
\end{figure}

\begin{figure}
\centering
\includegraphics[width=0.5\textwidth]{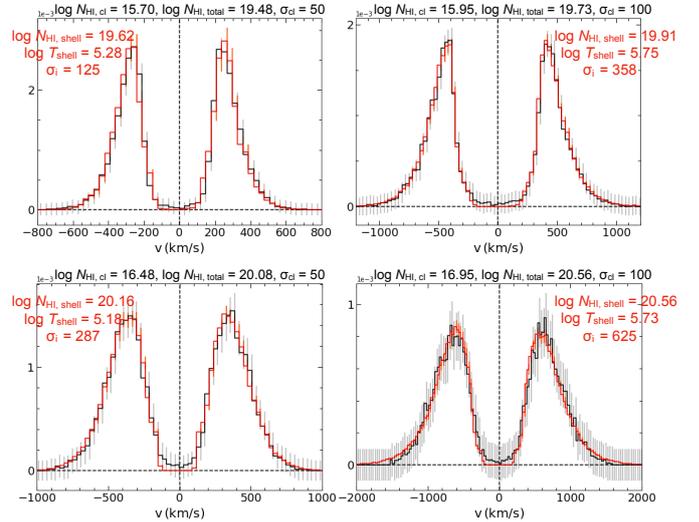}
    \caption{\textbf{Examples of shell model best-fits to randomly moving clumpy slab models.} Four panels represent four different ($N_{\rm HI,\,total}, \sigma_{\rm cl}$) cases. The black curves represent the outflowing clumpy slab model spectra and the red curves represent the shell model best-fits. $T_{\rm shell}$ have been obtained at the expected values from Eq. (\ref{eq:Teff}) within uncertainties.
    \label{fig:sigma_fits}}
\end{figure}

\begin{figure}
\centering
\includegraphics[width=0.45\textwidth]{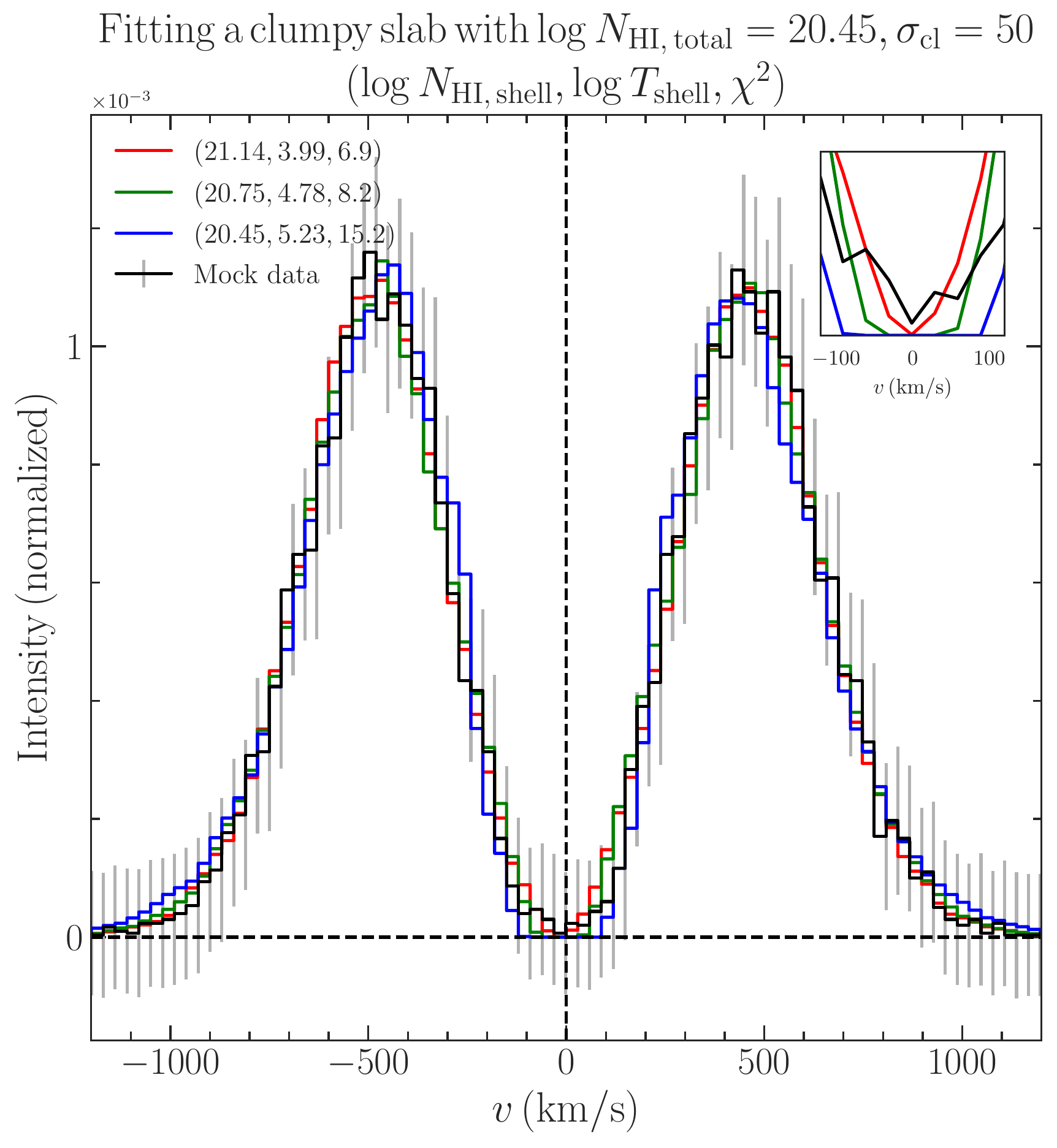}
    \caption{\textbf{Model examples showing the degeneracy between $N_{\rm HI,\,shell}$ and $T_{\rm shell}$.} The curves with different colors represent the clumpy slab model fitted (black) and the degenerate shell models with different $({\rm log}\,N_{\rm HI,\,shell}, {\rm log}\,T_{\rm shell}, {\rm \chi^{\rm 2}})$, obtained by fitting within a certain parameter subspace. As shown in the inset, the models with higher $T_{\rm shell}$ have more extended troughs at line center and thus are less favored in the fitting to the clumpy slab models (the mock data), which have sharper troughs at high $N_{\rm HI,\,total}$. 
    \label{fig:chi2}}
\end{figure}

\begin{figure}
\centering
\includegraphics[width=0.45\textwidth]{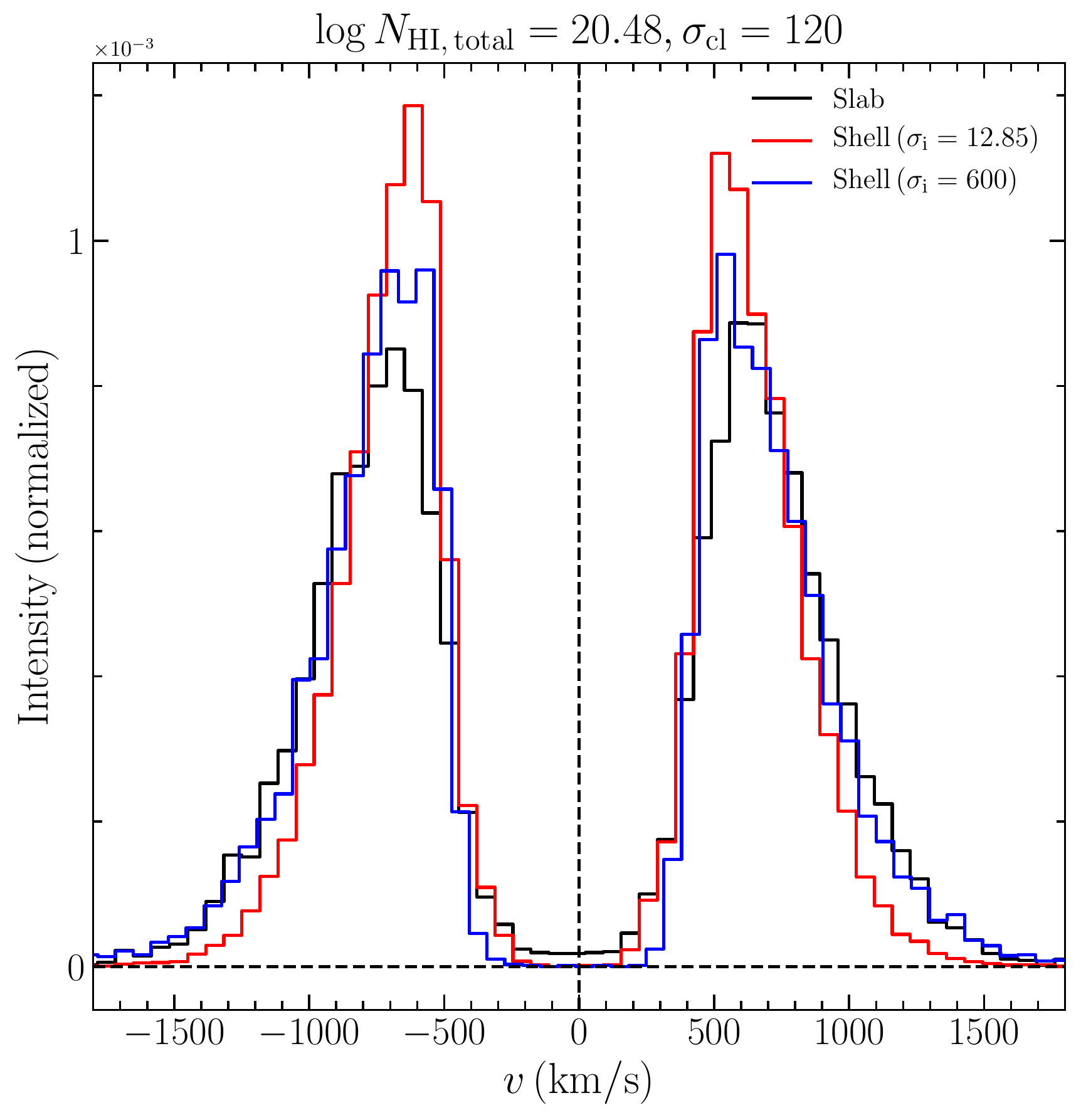}
    \caption{\textbf{Comparison between a clumpy slab model and two shell models with corresponding parameters and $\sigma_{\rm i}$ = 12.85 / 600\,km\,s$^{-1}$.} The black curve is an example clumpy slab model with $N_{\rm HI,\,total}$ and $\sigma_{\rm cl}$ values labeled on the top of the plot; the red and blue curves are the shell models at the expected $N_{\rm HI,\,total}$ and $T_{\rm shell}$ values with $\sigma_{\rm i}$ = 12.85\,km\,s$^{-1}$ and 600\,km\,s$^{-1}$, respectively. It is clear that the clumpy slab model tends to have lower peaks and larger fluxes near the line center than the corresponding homogeneous shell model with a small $\sigma_{\rm i}$; such a mismatch is mitigated by the broadening effect of a large $\sigma_{\rm i}$.
    \label{fig:small_sigmai}}
\end{figure}

We further add a random velocity dispersion (a Gaussian with standard deviation of $\sigma_{\rm cl}$ for all three dimensions) to the clumps and attempt to correlate it with certain shell model parameters, such as the internal random motion (or the effective temperature of the shell, $T_{\rm shell}$) of the shell model and the line width of the intrinsic \lya\ emission. We fit the clumpy slab model spectra with the large shell model grid, and the parameter values of the mock data are given in Table \ref{tab:parameter}. 

As shown in Figure \ref{fig:sigma}, we find that:

\begin{enumerate}

\item The derived $N_{\rm HI,\,shell}$ values are around the $N_{\rm HI,\,total}$ values, but a noticeable deviation has emerged. On average, $N_{\rm HI,\,shell}$ tends to be systemically higher than $N_{\rm HI,\,total}$ by $\sim 0.15$ dex (a factor of 1.5), especially at $N_{\rm HI,\,total} >\,10^{20} \rm cm^{-2}$;

\item The shell effective temperatures ($T_{\rm shell}$) are obtained at the effective temperatures of the clumpy slab model, defined as:
\begin{equation}
T_{\rm eff,\,slab} = T_{\rm cl} + \frac{\sigma_{\rm cl}^2 m_{\rm H}}{2 k_{\rm B}}
\label{eq:Teff}
\end{equation}
where $T_{\rm cl}$ is the kinematic temperature of one clump (fixed to 10$^{4}$\,K), $m_{\rm H}$ is the hydrogen atom mass and $k_{\rm B}$ is the Boltzmann constant. As the maximum $T_{\rm shell}$ of our large shell model grid is set to be 10$^{5.8}$\,K, we only explore $\sigma_{\rm cl}$ up to $\sim$ 100\,km\,s$^{-1}$, but we have verified that a larger $\sigma_{\rm cl}$ would still correspond to a $T_{\rm shell}$ value given by Eq. (\ref{eq:Teff});

\item Large $\sigma_{\rm i}$ values (several times of $\sigma_{\rm cl}$) are required to reproduce the wings of the clumpy slab models. These $\sigma_{\rm i}$ values are also positively correlated with $\sigma_{\rm cl}$ and $N_{\rm HI,\,shell}$, as shown in the bottom panel of Figure \ref{fig:sigma}. 
\end{enumerate}

We show four examples of shell model best-fits to the clumpy slab models in Figure \ref{fig:sigma_fits}. We find that (i) is due to the $N_{\rm HI,\,shell} \propto T_{\rm shell}^{-0.5}$ degeneracy. As we have detailed in Section \ref{sec:degeneracy}, in the optically thick regime where $a \tau_{0} = 2.8 \times 10^{-13} (N_{\rm HI,\,shell}/T_{\rm shell}) \gtrsim 10^3$, shell models with the same $N_{\rm HI,\,shell} T_{\rm shell}^{0.5}$ have almost identical line profiles, except that the ones with higher $T_{\rm shell}$ have slightly more extended troughs at the line center\footnote{This is because the cross section function of a higher $T_{\rm shell}$ is more extended near the line center (see Eq. (54) and (55) in \citealt{Dijkstra17}).}. This explains the deviation of $N_{\rm HI,\,shell}$ towards higher values at high \HI\ column densities, where the trough of the clumpy slab model becomes ``sharper'' as the flux density at line center approaches zero, and is better fitted at a slightly lower $T_{\rm shell}$ (and hence higher $N_{\rm HI,\,shell}$). We illustrate this effect in Figure \ref{fig:chi2}. In other words, $N_{\rm HI,\,shell}$ is still consistent with $N_{\rm HI,\,total}$ if we account for this $N_{\rm HI,\,shell} \propto T_{\rm shell}^{-0.5}$ degeneracy.

Moreover, (iii) is due to the intrinsic differences between the clumpy slab model and the shell model. As shown in Figure \ref{fig:small_sigmai} (cf. Figure 5 in \citealt{Gronke17b}), for $\sigma_{\rm cl} > $ 0, the clumpy slab model (black curve) tends to have lower peaks and larger fluxes near the line center, as compared to the corresponding homogeneous shell model (red curve). Therefore, in order to obtain a good fit, a large $\sigma_{\rm i}$ is required to flatten the peaks and spread the fluxes out into the wings (blue curve). As $\sigma_{\rm cl}$ or $N_{\rm HI,\,total}$ increases, the difference between the peak fluxes of two different models becomes larger, which requires a larger $\sigma_{\rm i}$. If we force $\sigma_{\rm i}$ to be small, the shell models would fail to fit the clumpy slab model, as a much higher ${\rm H\,{\textsc {i}}}$ column density than $N_{\rm HI,\,total}$ is required to fit the broad wings and it will inevitably yield a significant mismatch in the peaks.

As large $\sigma_{\rm i}$ values have been shown to be inconsistent with the observed nebular emission line widths (e.g. \ha\ or \hb, \citealt{Orlitova18}), it is reasonable to postulate that the clumpy model is a more realistic description of the actual gas distribution in ISM/CGM, as it naturally alleviates such discrepancies with moderate velocity dispersions of the clumps (see also \citealt{Li21b}). We will further discuss this point in \S\ref{sec:Interpretation}. 

\subsection{Clumpy Slab with Outflowing Clumps}\label{sec:outflow}

\begin{figure}
\centering
\includegraphics[width=0.47\textwidth]{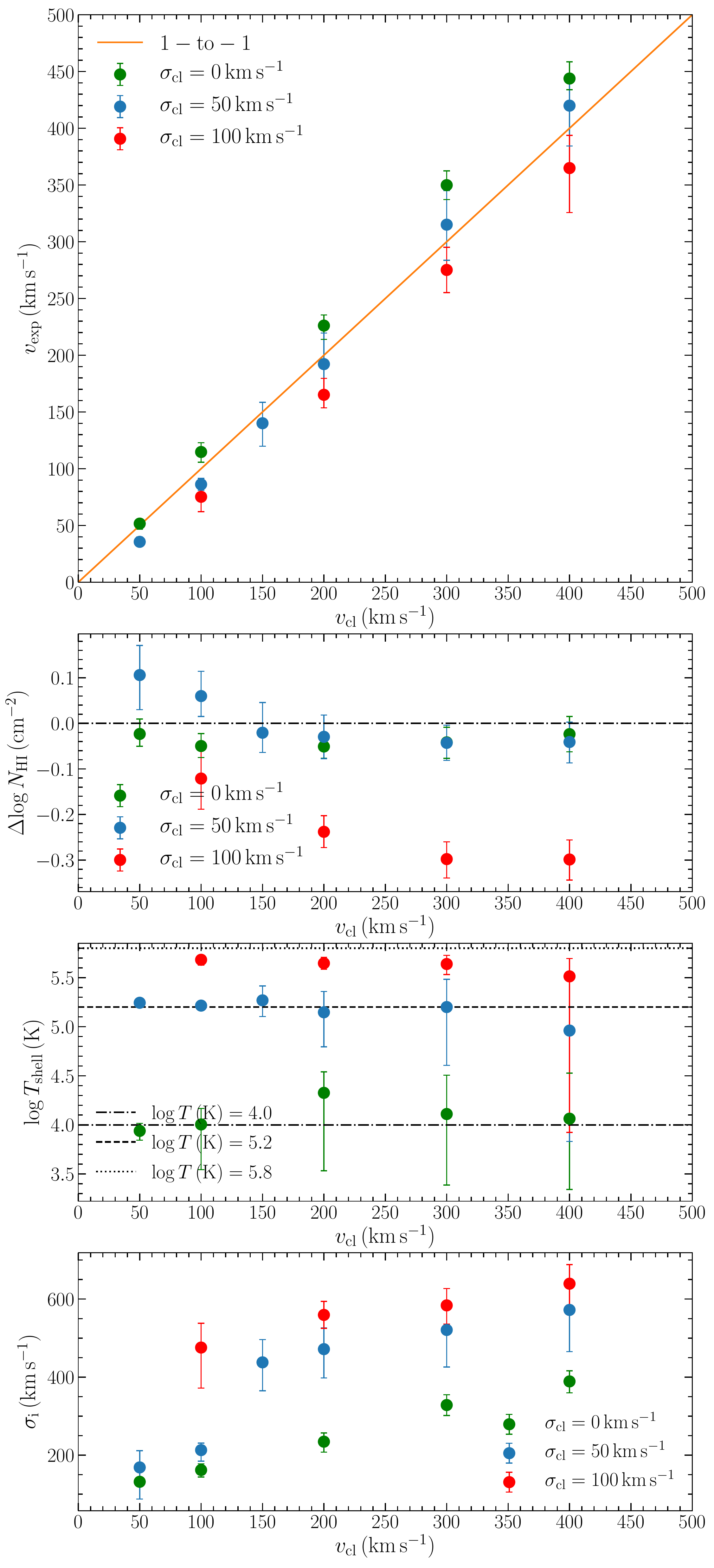}
    \caption{\textbf{Results of fitting clumpy slab models with outflowing moving clumps with shell models.} \emph{Upper:} The derived shell expansion velocities are mostly at the clump outflow velocities; \emph{Upper Middle:} $N_{\rm HI,shell}$ are reproduced mostly at $N_{\rm HI,total}$, albeit with several outliers with $\Delta {\rm log}\,N_{\rm HI} \gtrsim$ 0.3 dex; \emph{Lower Middle:} The derived shell temperatures are mostly at the effective temperatures of the clumpy slab model; \emph{Lower:} The distribution of the derived intrinsic \lya\ line widths ($\sigma_{\rm i}$) of the best-fit shell models, which increase as $\sigma_{\rm cl}$ or $v_{\rm cl}$ increases. The green, blue and red points represent the $\sigma_{\rm cl} =$ 0, 50 and 100 $\rm km\,s^{-1}$ models, respectively.
    \label{fig:vcl}}
\end{figure}

We further attempt to add a uniform outflow velocity ($v_{\rm cl}$) to the clumps and correlate it with the shell expansion velocity ($v_{\rm exp}$). We consider two different cases: (1) $v_{\rm cl} > 0$, $\sigma_{\rm cl} = 0$; (2) $v_{\rm cl} > 0$, $\sigma_{\rm cl} > 0$ , i.e. outflowing clumps without and with clump random motion, respectively. We find that in both cases, a considerably large $\sigma_{\rm i}$ is still required to achieve decent fits, otherwise the shell model best-fit would have a dip between two peaks on the red side, whereas the clumpy slab model has only one smooth red peak. 

\begin{figure}
\centering
\includegraphics[width=0.5\textwidth]{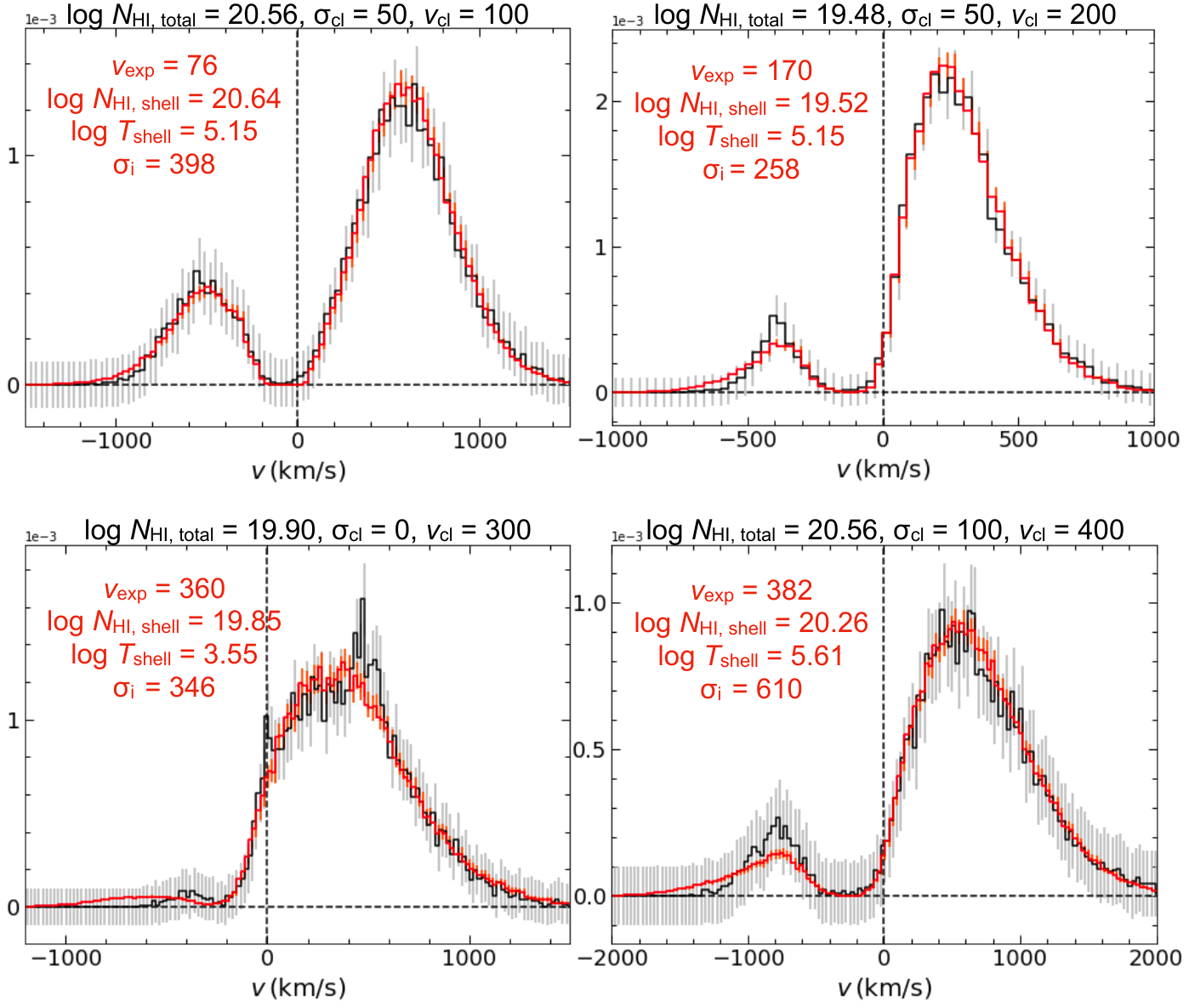}
    \caption{\textbf{Examples of outflowing shell model best-fits to outflowing clumpy slab models.} Four panels represent four different ($N_{\rm HI,\,total}, \sigma_{\rm cl}, v_{\rm cl}$) cases. The black curves represent the outflowing clumpy slab model spectra and the red curves represent the shell model best-fits. $v_{\rm exp}$ have been obtained at the expected values within uncertainties.
    \label{fig:vcl_fits}}
\end{figure}

As shown in Figure \ref{fig:vcl}, we find that:

\begin{enumerate}
    \item The fitted $v_{\rm exp}$ values are mostly consistent with $v_{\rm cl}$ within uncertainties;
    \item $N_{\rm HI,shell}$ are mostly reproduced at $N_{\rm HI,total}$, albeit with several outliers with $\Delta {\rm log}\,N_{\rm HI} \gtrsim$ 0.3 dex in the large $\sigma_{\rm cl}$ cases; 
    \item $T_{\rm shell}$ are mostly reproduced at $T_{\rm eff,slab}$ within uncertainties;
    \item Large $\sigma_{\rm i}$ values are still required and they increase as $\sigma_{\rm cl}$ or $v_{\rm cl}$ increases.
\end{enumerate}
We show four examples of outflowing shell model best-fits to outflowing clumpy slab models in Figure \ref{fig:vcl_fits} to illustrate the quality of the fits.

\begin{figure}
\centering
\includegraphics[width=0.5\textwidth]{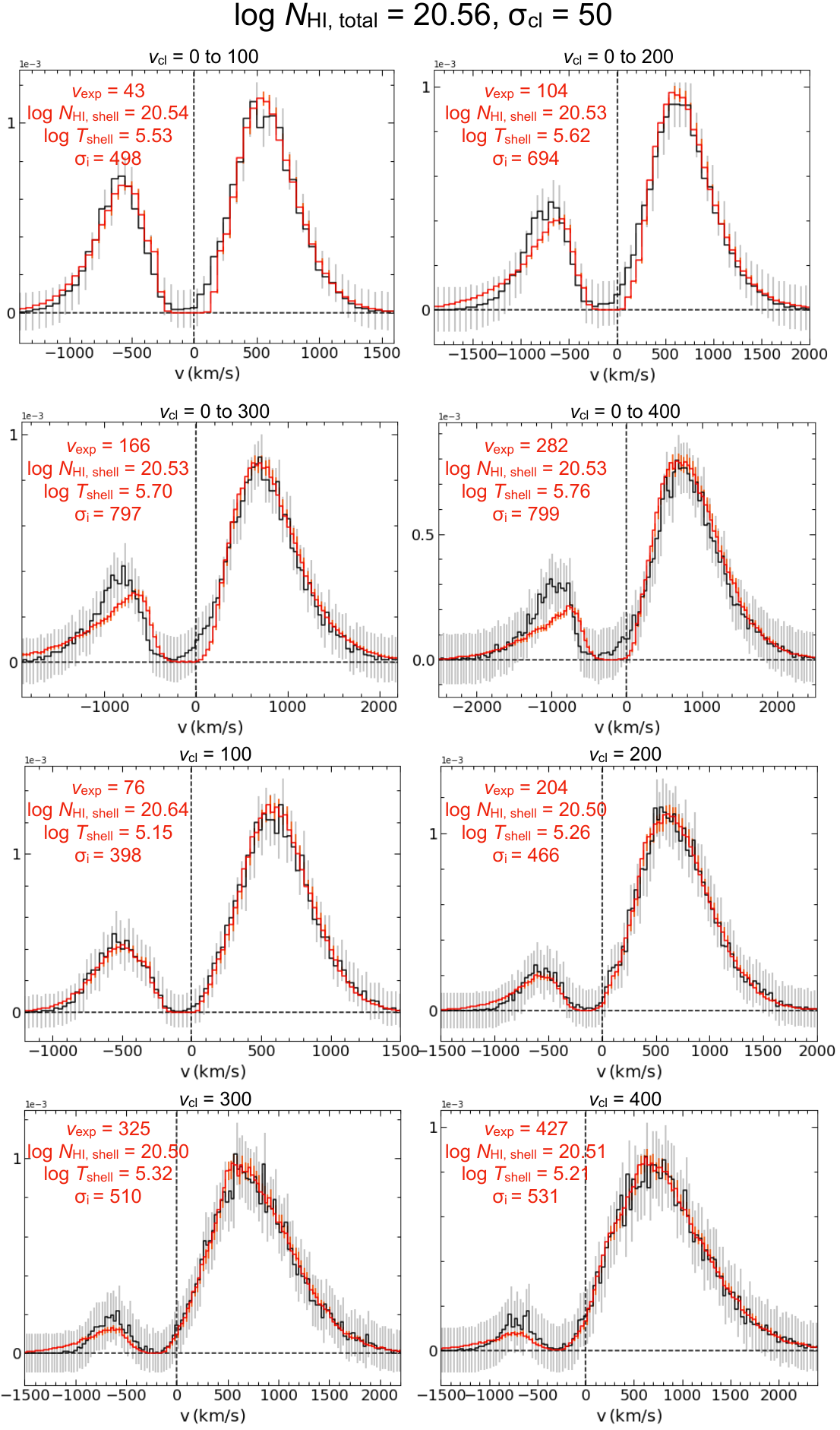}
    \caption{\textbf{Examples of clumpy slab models with linearly increasing outflow velocities, compared to those with constant outflow velocities.} The upper two rows are four models with linearly increasing outflow velocities, whereas the lower two rows are four models with constant outflow velocities. For the models with linearly increasing outflow velocities, the red-to-blue peak flux ratio is much lower. The peak separation is also larger, which boosts the fitted $T_{\rm shell}$ and $\sigma_{\rm i}$ values, but $N_{\rm HI,\,shell} \simeq N_{\rm HI,\,total}$ remains true. $v_{\rm exp}$ is roughly obtained at $\frac{1}{2}v_{\rm max}$.
    \label{fig:linearv}}
\end{figure} 

\begin{figure}
\centering
\includegraphics[width=0.45\textwidth]{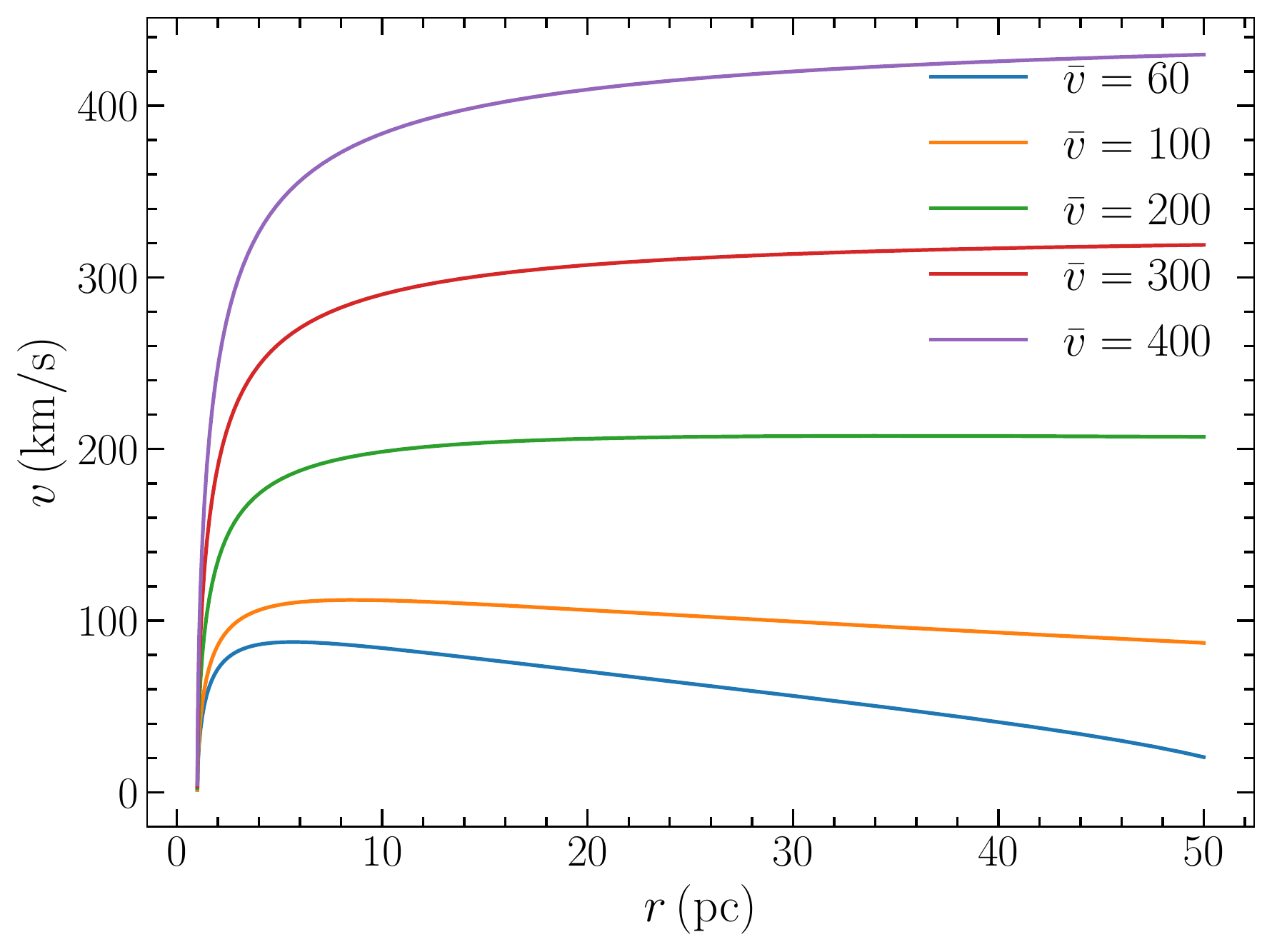}
    \caption{\textbf{Examples of momentum-driven radial velocity profiles given by Eq. (\ref{eq:solution})}. Different colored curves represent five $v(r)$ profiles with different ($\sigma_{\rm cl}$, $v_{\rm cl}$). Here $\sigma_{\rm cl}$ is fixed to $50\,\rm km\,s^{-1}$ and $v_{\rm cl,\,\infty}$ are adjusted to make the average radial velocity, $\overline{v(r)} = 60, 100, 200, 300$ and $400\,\rm km\,s^{-1}$. The acceleration decreases with radius, and the velocity either flattens or drops at large $r$, depending on the actual values of $\sigma_{\rm cl}$ and $v_{\rm cl}$.
    \label{fig:v_profiles}}
\end{figure} 

\begin{figure}
\centering
\includegraphics[width=0.45\textwidth]{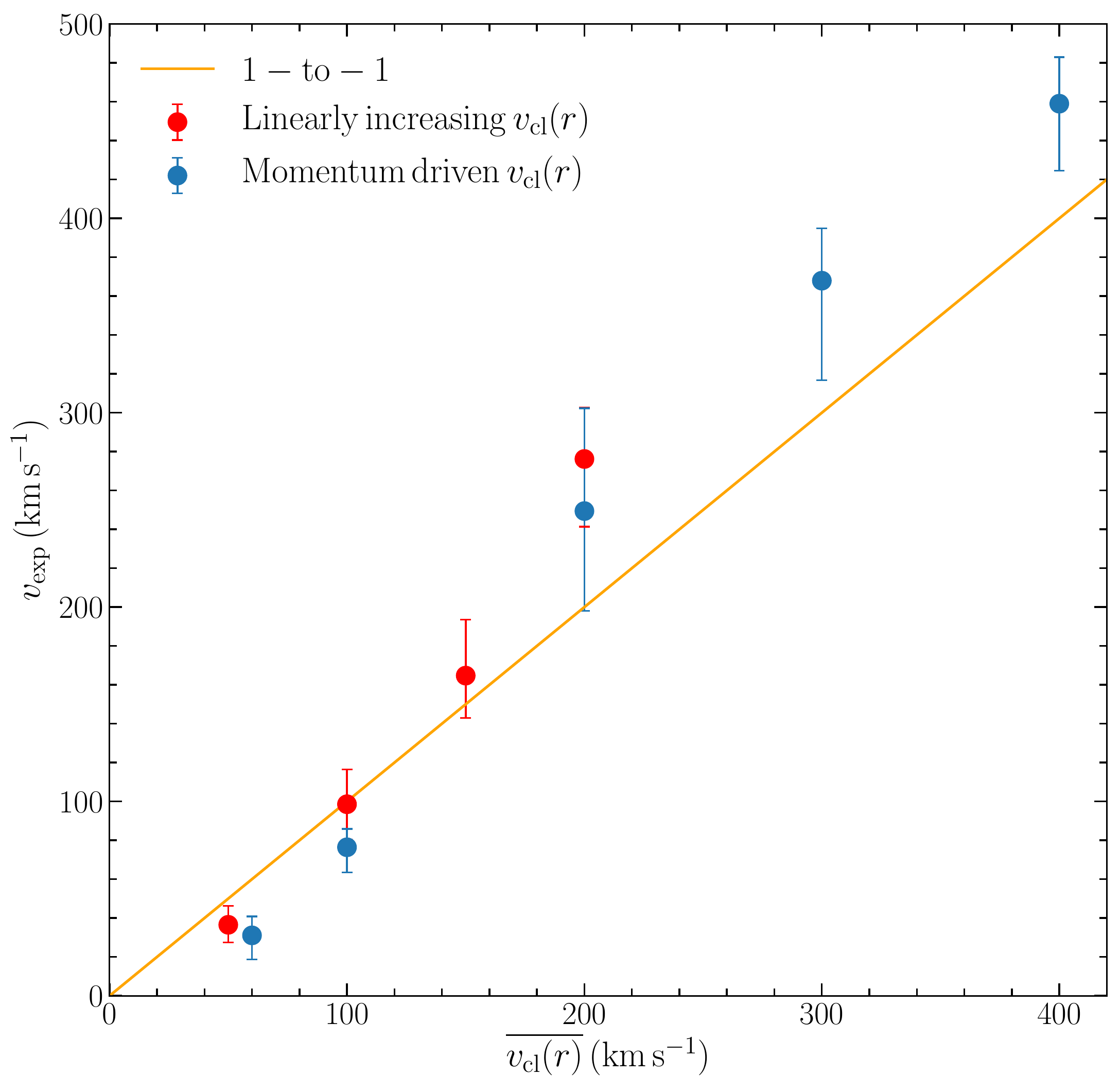}
    \caption{\textbf{Relation between the average radial velocity $\overline{v_{\rm cl}(r)}$ of clumpy slab models and the derived $v_{\rm exp}$ from shell model fitting.} The red points represent the linearly increasing scenario, and the blue points represent the momentum driven scenario. For both scenarios, $\overline{v_{\rm cl}(r)}$ and $v_{\rm exp}$ are basically consistent, suggesting that shell model fitting probes the average radial velocity of the clumps.
    \label{fig:radial_v}}
\end{figure} 

In addition to adding a uniform clump outflow velocity, we have also experimented with two more sophisticated velocity profiles. Firstly, we add a Hubble flow-like outflow velocity increasing linearly from 0 (at the center of the simulation region) to $v_{\rm max}$ (at the boundary of the simulation region) to the clumps and fit the \lya\ model spectra. The results are shown in Figure \ref{fig:linearv}. The four panels in the first two rows represent four models with linearly increasing outflow velocities, as compared to the lower four panels with constant outflow velocities. It can be seen from the line profiles that for the models with linear increasing outflow velocities: (1) the red-to-blue peak flux ratio is much lower than that of the corresponding uniform outflow model (either with $v_{\rm cl} = \frac{1}{2}v_{\rm max}$ or $v_{\rm max}$), which is not well captured by the best-fit shell model, especially at high outflow velocities; (2) the peak separation is also larger than that of the corresponding uniform outflow model. As a result, the fitted $T_{\rm shell}$ and $\sigma_{\rm i}$ values have been boosted due to (2), but $N_{\rm HI,\,shell} \simeq N_{\rm HI,\,total}$ remains true. $v_{\rm exp}$ is roughly obtained at $\frac{1}{2}v_{\rm max}$, but should be considered as an upper limit as it actually yields a best-fit spectrum with a red-to-blue peak flux ratio higher than that of the fitted clumpy model.

Secondly, we consider a scenario where the clumps are accelerated in a momentum-driven manner and in the meantime, decelerated by a gravitational force. This is motivated by the fact that in real galactic environments, the cool clouds can be accelerated by radiation pressure or ram pressure of the hot wind as they break out of the ISM, as they are decelerating within the gravitational well of the dark matter halo. The momentum equation of a clump can be then written as \citep{Murray05, Dijkstra12}:

\begin{equation}
\frac{\mathrm{d}v(r)}{\mathrm{d}t}=-\frac{GM(r)}{r^2}+Ar^{-\alpha}
\label{eq:momentum}
\end{equation}
where $r$ is the clump radial position, $v(r)$ is the clump radial outflow velocity, and $M(r)$ is the total mass within $r$. Here the clump acceleration is determined by two competing terms on the right hand side, the first of which is due to gravitational deceleration and the second of which is an empirical power-law acceleration term \citep{Steidel10}. Assuming the gravitational potential is of an isothermal sphere, then $M(r)=2 \sigma_{\rm cl}^2\,r/G$, where $\sigma_{\rm cl}$ is the velocity dispersion of the clumps. Eq. (\ref{eq:momentum}) can then be analytically solved as:

\begin{align}
v(r) = \sqrt{4\,\sigma_{\rm cl}^2\,{\rm ln}\Big{(}\frac{r_{\rm min}}{r}\Big{)} + v_{\rm cl,\,\infty}^2 \Big{(}1-\Big{(}\frac{r}{r_{\rm min}}\Big{)}^{1-\alpha}\Big{)}}
\label{eq:solution}
\end{align}
where $r_{\rm min}$ is the inner cutoff radius that satisfies $v(r_{\rm min})=0$, and $v_{\rm cl,\,\infty}$ = $\sqrt{2Ar_{\rm min}^{1-\alpha}/(\alpha-1)}$ is the asymptotic maximum outflow velocity if there were no gravitational deceleration. Following \citet{Dijkstra12}, we have fixed $r_{\rm min}$ = 1 pc (note the clumpy slab model has a half height of 50 pc and the model is re-scalable by design) and $\alpha$ = 1.4\footnote{These choices come from the clump radial velocity models that provide good fits to the observed \lya\ surface brightness profiles \citep{Dijkstra12}.} and left $\sigma_{\rm cl}$ and $v_{\rm cl}$ as free parameters. 

We show five examples of $v(r)$ with different ($\sigma_{\rm cl}$, $v_{\rm cl}$) in Figure \ref{fig:v_profiles}. It can be seen that the acceleration decreases with radius, and the velocity either flattens or drops at large $r$, depending on the actual values of $\sigma_{\rm cl}$ and $v_{\rm cl}$. We fix $\sigma_{\rm cl} = 50\,\rm km\,s^{-1}$ and adjust $v_{\rm cl,\,\infty}$ to achieve average radial velocities of $\overline{v_{\rm cl}(r)} = 60, 100, 200, 300$ and $400\,\rm km\,s^{-1}$. We then assign these radial velocity profiles to the clumps and fit their model spectra with shell models. We find that $N_{\rm HI,\,shell}$ and $T_{\rm shell}$ are reproduced at the expected values, and $v_{\rm exp}$ is roughly obtained at the average radial velocity, $\overline{v_{\rm cl}(r)}$.

We show that the derived $v_{\rm exp}$ is consistent with the average radial velocity $\overline{v_{\rm cl}(r)}$ for these two scenarios, as shown in Figure \ref{fig:radial_v}. In reality, if the velocity distribution of the \lya\ scattering clumps is semi-linear or similar to the ``momentum-driven + gravitational deceleration'' scenario, the shell model fitting will probe the average outflow velocity of the clumps. Assuming the same clumps are responsible for producing metal absorption (and ignoring effects due to an anisotropic gas distribution), the clump velocity distribution can be constrained by observations on UV absorption lines. One should then expect the outflow velocity $v_{\rm exp}$ output from shell model fitting to be consistent with the average of the absorption velocities. If the clump velocity distribution is non-linear, $v_{\rm exp}$ may no longer be an average outflow velocity, but should still lie between the minimum and maximum absorption velocities.

\subsection{Clumpy Slab with an Inter-Clump Medium (ICM)}\label{sec:multiphase_slab}

\begin{figure}
\centering
\includegraphics[width=0.47\textwidth]{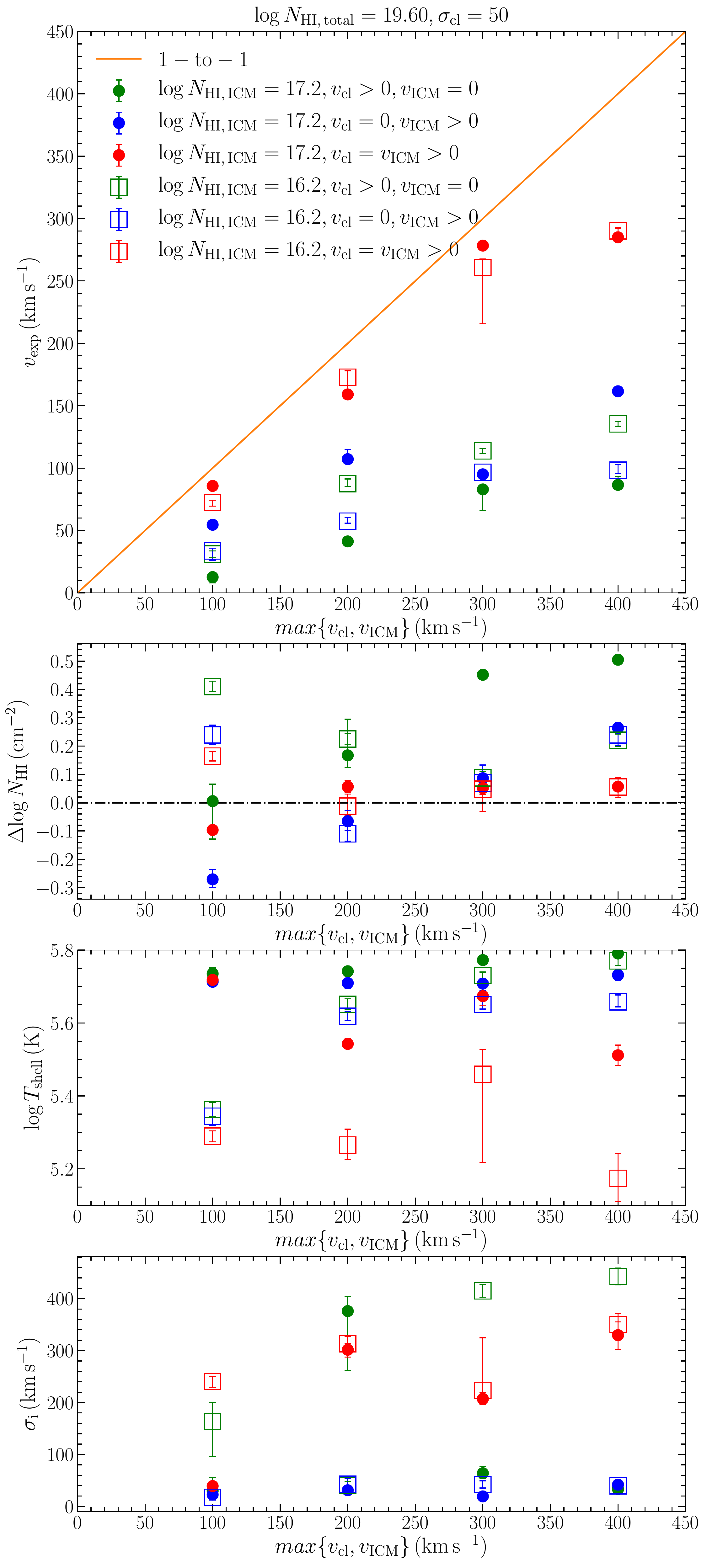}
    \caption{\textbf{Results of fitting clumpy slab models with static and outflowing ICM with shell models.} \emph{Upper:} The derived shell expansion velocities are much smaller than the clump/ICM outflow velocities unless both components are co-outflowing, which yields $v_{\rm exp} \simeq v_{\rm cl} = v_{\rm ICM}$; \emph{Upper Middle:} $N_{\rm HI,shell}$ are mostly at $N_{\rm HI,total}$, albeit with several outliers with $\Delta {\rm log}\,N_{\rm HI} \gtrsim$ 0.3 dex; \emph{Lower Middle:} The derived shell temperatures are boosted by the hot ICM to be higher than the effective temperatures of the clumpy slab model; \emph{Lower:} The distribution of the derived intrinsic \lya\ line widths ($\sigma_{\rm i}$) of the best-fit shell models, which increase as $\sigma_{\rm cl}$ or $v_{\rm cl}$ increases. The green, blue and red points represent (1) $v_{\rm cl} > 0, v_{\rm ICM} = 0$ (2) $v_{\rm cl} = 0, v_{\rm ICM} > 0$ (3) $v_{\rm cl} = v_{\rm ICM} > 0$ models, respectively.
    \label{fig:vICM}}
\end{figure}

Motivated by the fact that in real astrophysical environments (e.g. in the CGM; \citealp{Tumlinson17}), a hot, highly-ionized gas phase with residual \HI\ exists and affects \lya\ RT \citep{Laursen13}, we further add another hot phase of gas between the clumps to the clumpy slab model as the inter-clump medium (ICM). Although the total column density of the low-density ICM ($\sim n_{\rm HI,\,{\rm ICM}}r_{\rm gal} \lesssim 10^{\rm -4}\,\rm cm^{-3} \times kpc \sim 10^{\rm 17}\,\rm cm^{-2}$) is supposed to be several orders of magnitude lower than the typical values of $N_{\rm HI,\,total}$ from the cool clumps, it has several non-negligible effects on the \lya\ spectra. We find that adding another hot phase of gas at 10$^{6}$\,K (the typical temperature of diffuse gas in a dark matter halo) will: (1) deepen the trough at line center; (2) increase the peak separation; (3) modify the red-to-blue peak flux ratio of the model \lya\ spectrum\footnote{Regarding the effects of ICM with different temperatures and column densities, we refer the readers to Appendix \ref{sec:Ticm_spec} and \ref{sec:Ticm_fccrit}.}.

Here we consider two different scenarios: static ICM and outflowing ICM. For both scenarios, we generate two sets of multiphase, clumpy slab models with $n_{\rm HI,\,{\rm ICM}} = 10^{-4}$ and $10^{-3}\,\rm cm^{-3}$ (or equivalently, $N_{\rm HI,\,{\rm ICM}} = 10^{16.2}$ and $10^{17.2}\,\rm cm^{-2}$) and fit them using the large shell model grid. The values of the input and output parameters are shown in Figure \ref{fig:vICM}. We hereby discuss two scenarios respectively:

\subsubsection{Static hot ICM: $T_{\rm ICM} = 10^{6}$\,K, $v_{\rm ICM} = 0$}\label{sec:icm1}

We find that adding a static hot ICM increases the peak separation and decreases the red-to-blue peak flux ratio of the model \lya\ spectrum. These two effects can be seen by comparing the first and second rows of Figure \ref{fig:vICM_fits}. In terms of the shell model best-fit parameters, the former effect increases $T_{\rm shell}$ but does not boost $N_{\rm HI,\,total}$ significantly, and the latter effect decreases $v_{\rm exp}$ to $\ll v_{\rm cl}$. These effects are shown in Figure \ref{fig:vICM} by green circles and open squares (which correspond to two different ICM \HI\ column densities).

\subsubsection{Outflowing hot ICM: $T_{\rm ICM}$ = 10$^{6}$\,K, $v_{\rm ICM} > 0$}\label{sec:icm2}

\begin{figure*}
\centering
\includegraphics[width=\textwidth]{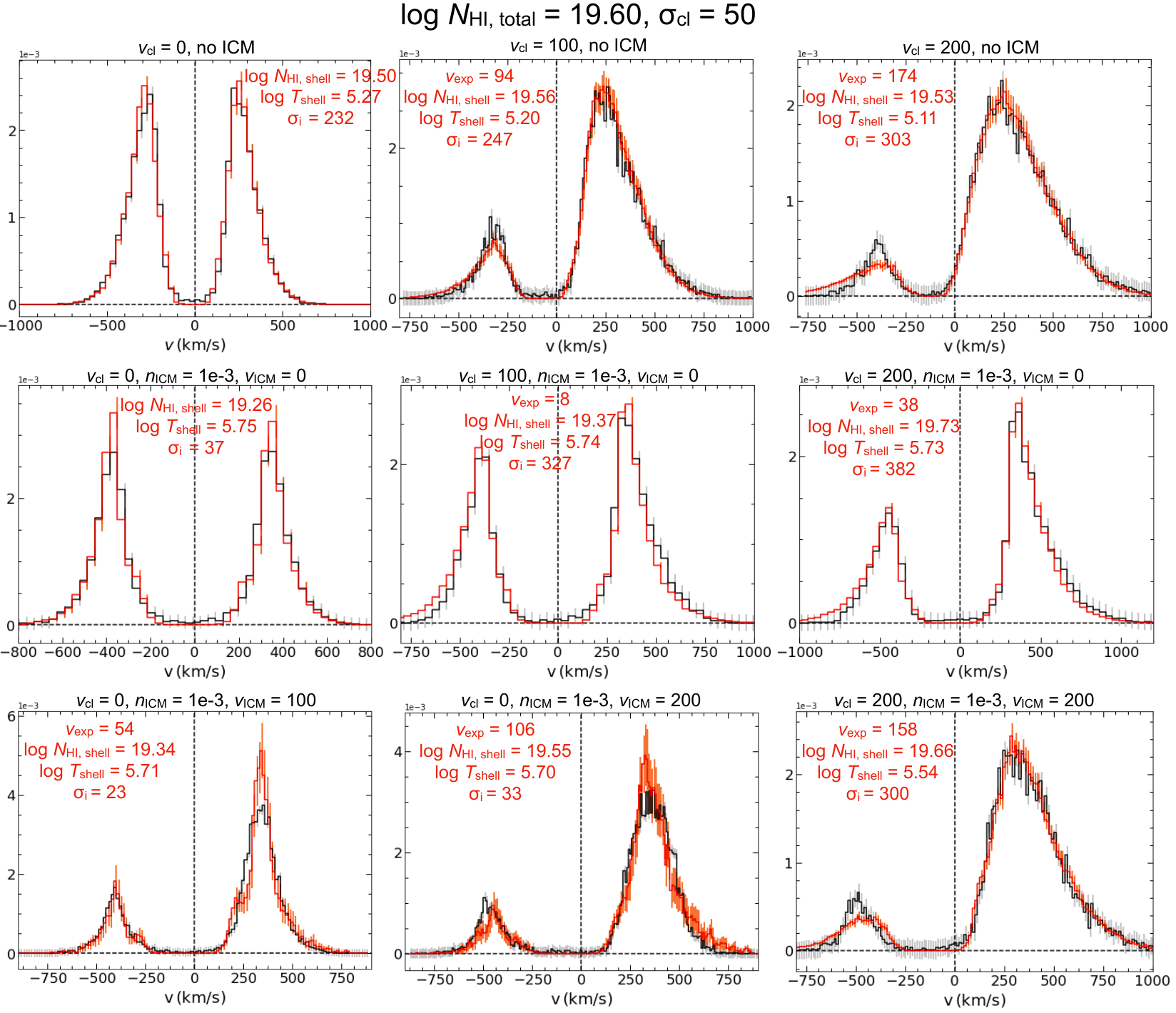}
    \caption{\textbf{Examples of shell model best-fits to outflowing clumpy slab models with and without ICM.}  The first row represents three $v_{\rm cl}$ > 0 cases without ICM. The second row represents three $v_{\rm cl}$ > 0 cases with a static, $T = 10^{6}\,\rm K$, $n_{\rm HI} = 10^{-3}\,\rm cm^{-3}$ ICM. Adding this hot phase of static ICM tends to: (1) deepen the trough at line center; (2) increase the peak separation; (3) decrease the red-to-blue peak flux ratio. The third row represents three $v_{\rm cl}$ > 0 cases with an outflowing ICM. A $v_{\rm ICM}$ > 0 ICM will further increase the red-to-blue peak flux ratio and increase the $v_{\rm exp}$ of the shell model best-fit. In particular, the model with the same clump and ICM outflow velocity prefers a shell expansion velocity $v_{\rm exp} \simeq v_{\rm cl} = v_{\rm ICM}$.
    \label{fig:vICM_fits}}
\end{figure*}

As shown in the third row of Figure \ref{fig:vICM_fits}, adding an outflow velocity to the ICM will increase the red-to-blue peak flux ratio. The first two panels have $v_{\rm cl}$ = 0 and $v_{\rm ICM}$ > 0, whereas the third panel has $v_{\rm cl}$ = $v_{\rm ICM}$ > 0. Notably, the quality of the best-fits has become worse, suggesting the non-linear effect of a hot ICM on the \lya\ model spectra. The $T_{\rm shell}$ and $N_{\rm HI,\,total}$ values are similar to the static ICM case. Interestingly, in the first two panels where $v_{\rm cl} = 0$ and $v_{\rm ICM} > 0$, we have $v_{\rm exp} \simeq \frac{1}{2} v_{\rm ICM}$; whereas in the third panel where $v_{\rm cl} = v_{\rm ICM}$ > 0, the red-to-blue peak flux ratio becomes similar to the no-ICM case (the third panel in the first row), and $v_{\rm exp} \simeq v_{\rm cl} = v_{\rm ICM}$ is obtained.

It is therefore evident that if a multiphase, clumpy slab model with ($v_{\rm cl}$, $v_{\rm ICM}$) and a shell model with $v_{\rm exp}$ give the same \lya\ spectrum (especially the same red-to-blue peak flux ratio), then $v_{\rm exp}$ should lie between $v_{\rm cl}$ and $v_{\rm ICM}$. In particular, if $v_{\rm cl} = v_{\rm ICM}$, i.e., the cool clumps and the hot ICM are co-outflowing at the same speed, we would expect $v_{\rm exp} = v_{\rm cl} = v_{\rm ICM}$. In reality, we expect the cool clumps to be entrained by the local flow of hot gas (i.e. $v_{\rm cl} \simeq v_{\rm ICM}$), as a large velocity difference between two phases of gas may destroy the cool clumps quickly via hydrodynamic instabilities\footnote{Note that in a relatively rare 
scenario (e.g. very close to the launching radius of a galactic wind), the hot phase may be moving faster than the cool clumps, i.e., $v_{\rm ICM} > v_{\rm cl}$, and the shell model fitting would obtain $v_{\rm ICM} > v_{\rm exp} > v_{\rm cl}$.} (see e.g. \citealt{Klein94}). Therefore, we conclude that the gas outflow velocities extracted from fitting \lya\ spectra should be consistent between the shell model and the multiphase, clumpy model.

\subsection{Multiphase Clumpy Sphere}\label{sec:multiphase_sphere}
We further consider a more physically realistic gas geometric distribution, i.e., a multiphase clumpy sphere, which we have adopted in fitting observed spatially-resolved \lya\ spectra in \citet{Li21} and \citet{Li21b}. As a multiphase sphere model has an upper limit for the clump volume filling factor $F_{V}$ ($\lesssim$\,0.7 for numerical reasons) and hence for the covering factor $f_{\rm cl}$ (= 3$r_{\rm gal}$/4$r_{\rm cl}$\,$F_{\rm V}$ = 150\,$F_{\rm V}$), it cannot have an as high covering factor as the multiphase clumpy slab model (i.e. `less clumpy'). However, as long as $f_{\rm cl}$ is much larger than a critical value $f_{\rm cl,\,crit}$, the clumpy model would be sufficiently similar to a homogeneous model \citep{Gronke17b}. If the condition $f_{\rm cl} \gg f_{\rm cl,\,crit}$ is not satisfied, a considerable number of \lya\ photons would escape near the line center and yields residual fluxes at line center in the emergent spectrum. Here we explore the connection between multiphase, clumpy spherical models and shell models in these two physical regimes respectively: $f_{\rm cl} \gg f_{\rm cl,\,crit}$ and $f_{\rm cl} \simeq f_{\rm cl,\,crit}$.

\subsubsection{Very Clumpy Sphere: $f_{\rm cl} \gg f_{\rm cl,\,crit}$}\label{sec:sphere1}

\begin{figure*}
\centering
\includegraphics[width=\textwidth]{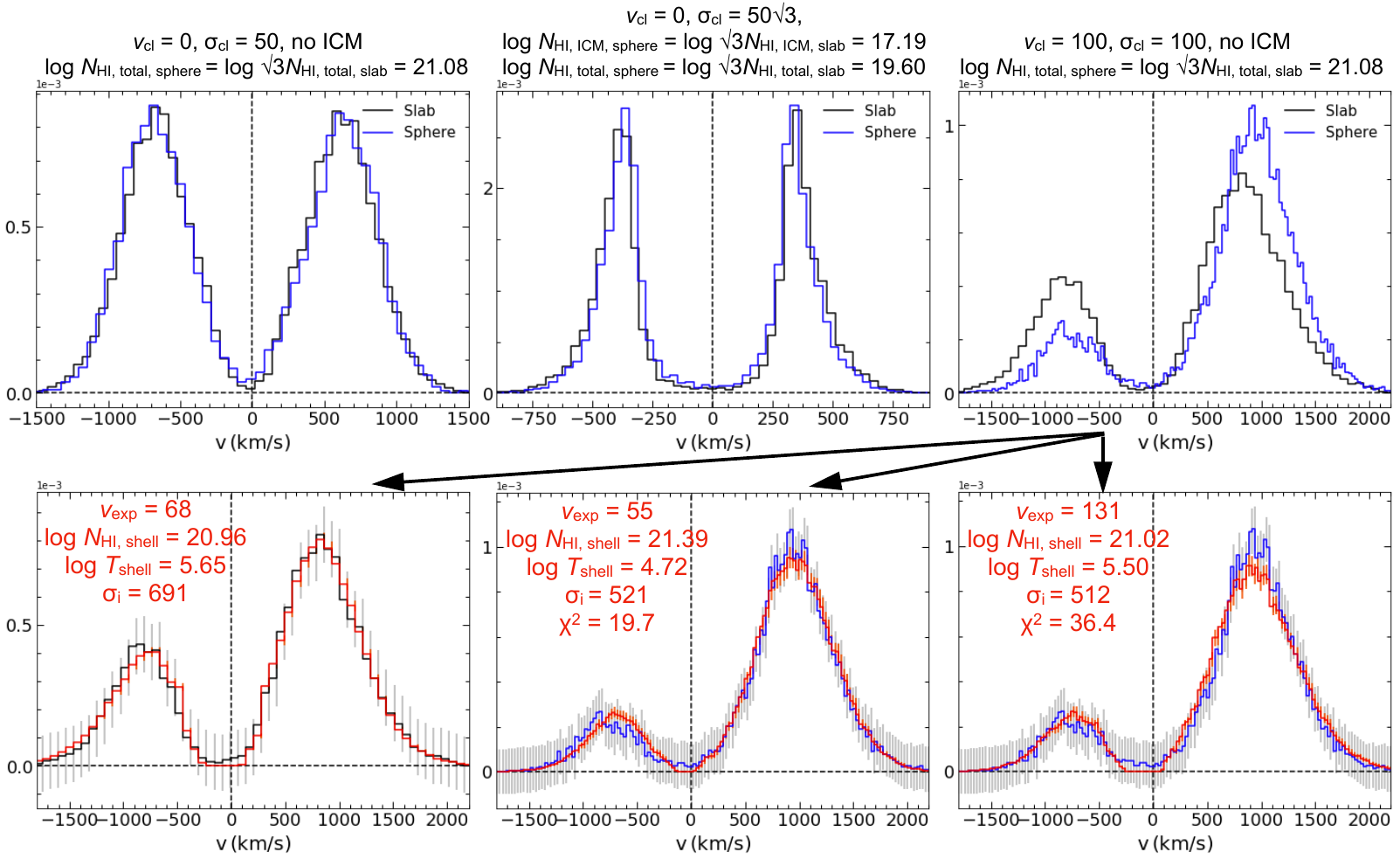}
    \caption{\textbf{Comparison between clumpy sphere models and clumpy slab models.} \emph{Upper row:} The first two panels show that a non-outflowing ($v_{\rm cl} = v_{\rm ICM}$ = 0) spherical model with ($\sigma_{\rm cl}, \sqrt{3}N_{\rm HI,\,total}, \sqrt{3}N_{\rm HI,\,ICM}$) gives an identical spectrum to a slab model with ($\sigma_{\rm cl}, N_{\rm HI,\,total}, N_{\rm HI,\,ICM}$), and hence yields the same shell model best-fit parameters. The factor $\sqrt{3}$ should arise from the geometrical difference a sphere and a slab (see \S\ref{sec:sphere1} for details). The third panel shows that adding the same $v_{\rm cl}$ yields a mismatch between the two models, which should be a non-linear effect due to the geometrical difference. \emph{Lower row:} The first panel shows the shell model best-fit to a clumpy slab model (with $v_{\rm exp} \simeq v_{\rm cl}$, $N_{\rm HI,\,shell} \simeq N_{\rm HI,\,total}$ and $T_{\rm shell} \simeq T_{\rm eff,\,slab}$), and the second panel shows the best-fit to the corresponding clumpy spherical model, where $v_{\rm exp}$ and $T_{\rm shell}$ are lower than expected and $N_{\rm HI,\,shell}$ is higher than expected. The third panel shows that if we restrict $T_{\rm shell}$ to be $\geq 10^{5.5}$\,K, the best-fit $v_{\rm exp}$ and $N_{\rm HI,\,shell}$ become closer to the expected values, although the best-fit gives a higher $\chi^2$ due to the mismatch in the red peak.
    \label{fig:sphere}}
\end{figure*}

For a very clumpy spherical model, i.e.  $f_{\rm cl} \gg f_{\rm cl,\,crit}$, we find that a non-outflowing ($v_{\rm cl} = v_{\rm ICM}$ = 0) spherical model with ($\sigma_{\rm cl}, \sqrt{3}N_{\rm HI,\,total}, \sqrt{3}N_{\rm HI,\,ICM}$) gives an identical spectrum to a slab model with ($\sigma_{\rm cl}, N_{\rm HI,\,total}, N_{\rm HI,\,ICM}$), and hence yields the same shell model best-fit parameters. This is shown in the first two panels in the first row of Figure \ref{fig:sphere}. The factor $\sqrt{3}$ should arise from the geometrical difference a sphere and a slab. Specifically, in the optically thick and $f_{\rm cl} \gg f_{\rm cl,\,crit}$ regime, the mean path length of \lya\ photons is $\sqrt{3}B$ for a slab and $R$ for a sphere, where $B$ is the slab half-height and $R$ is the sphere radius \citep{Adams75}. 

However, adding two different models the same outflow velocity to either the clumps or the ICM yields a mismatch, as shown in the third panel in the first row of Figure \ref{fig:sphere}. Such a mismatch should be due to the geometrical difference as well, yet we are unable to relate the two models with a scale factor in their outflow velocities (e.g. a spherical model with ($\sigma_{\rm cl}, \sqrt{3}N_{\rm HI,\,total}, \sqrt{3}v_{\rm cl}$) is still different from a slab model with ($\sigma_{\rm cl}, N_{\rm HI,\,total}, v_{\rm cl}$)). Therefore, we speculate that the geometrical difference has a non-linear effect on the propagation of the \lya\ photons through the outflowing \HI\ gas.

Nevertheless, we attempt to fit an outflowing clumpy spherical model with the shell model grid. The results are shown in the second row of Figure \ref{fig:sphere}. The first panel shows the shell model best-fit to a clumpy slab model (with $v_{\rm exp} \simeq v_{\rm cl}$, $N_{\rm HI,\,shell} \simeq N_{\rm HI,\,total}$ and $T_{\rm shell} \simeq T_{\rm eff,\,slab}$ as expected), and the second panel shows the best-fit to the corresponding clumpy spherical model, where $v_{\rm exp}$ and $T_{\rm shell}$ are lower than expected and $N_{\rm HI,\,shell}$ is higher than expected. However, we find that such a mismatch is due to the intrinsic parameter degeneracy of the shell model (see \S\ref{sec:outflow_deg}). If we restrict $T_{\rm shell}$ to be $\geq 10^{5.5}$\,K, the best-fit $v_{\rm exp}$ and $N_{\rm HI,\,shell}$ become consistent with the expected values, although the best-fit gives a higher $\chi^2$ due to a larger mismatch in the trough and red peak (as shown in the third panel). We therefore speculate that the correspondence between shell models and clumpy slab models still roughly holds for clumpy spherical models, with slightly larger uncertainties due to the inessential geometrical difference. 

\subsubsection{Moderately Clumpy Sphere: $f_{\rm cl} \simeq f_{\rm cl,\,crit}$}\label{sec:sphere2}
\begin{figure*}
\centering
\includegraphics[width=\textwidth]{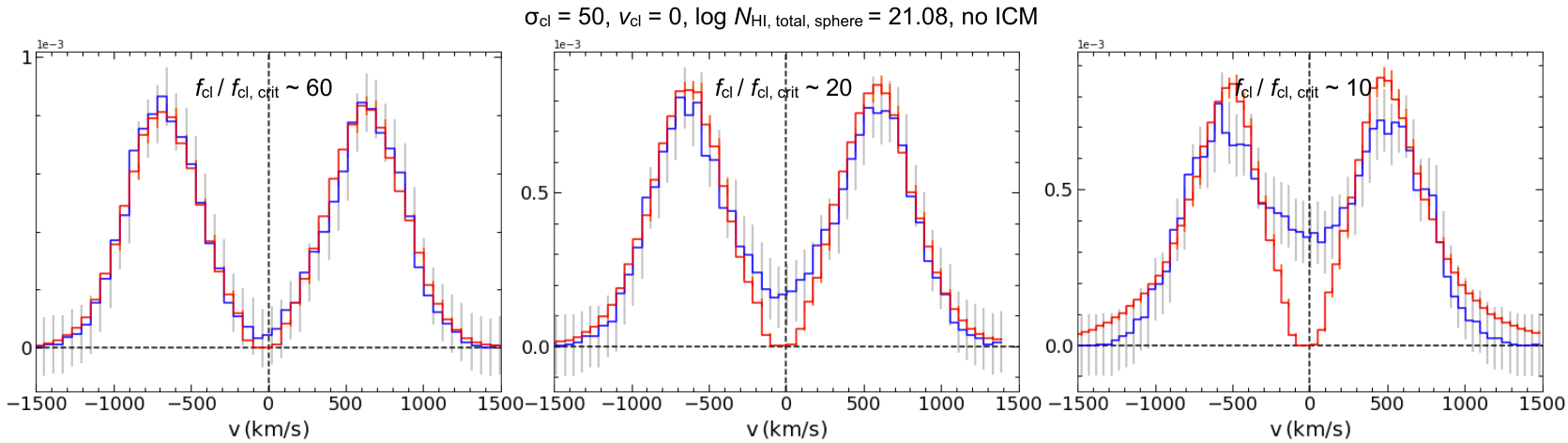}
    \caption{\textbf{Shell model fits to clumpy spherical models with decreasing $f_{\rm cl}/f_{\rm cl,\,crit}$.} The three panels correspond to three ($\sigma_{\rm cl}, v_{\rm cl}, {\rm log}\,N_{\rm HI,\,total}$) = (50, 0, 21.08) models with different $f_{\rm cl}/f_{\rm cl,\,crit}$ values: $\sim$ 60, 20, and 10, respectively. The blue curves are the clumpy spherical model spectra and the red curves are the shell model best-fits. As $f_{\rm cl}/f_{\rm cl,\,crit}$ decreases, in the beginning the shell model is still able to produce a decent fit with reasonable parameters (albeit with the mismatch at line center), but eventually the fit fails at $f_{\rm cl}/f_{\rm cl,\,crit} \simeq$ 10.
    \label{fig:fcl}}
\end{figure*}

If the spherical model is only moderately clumpy, i.e. $f_{\rm cl} \simeq f_{\rm cl,\,crit}$, the \lya\ optical depth at line center will be low enough for photons to escape, which yields a non-zero residual flux density at line center. We find that as $f_{\rm cl}/f_{\rm cl,\,crit}$ decreases, in the beginning the shell model is still able to produce a decent fit with reasonable parameters (albeit with the mismatch at line center), but eventually the fit fails at $f_{\rm cl}/f_{\rm cl,\,crit} \simeq$ 10. We illustrate this result in Figure \ref{fig:fcl}. In general, in the $f_{\rm cl} \simeq f_{\rm cl,\,crit}$ regime, no direct correlation has been found between the shell and clumpy model parameters due to the efficient escape of \lya\ photons at line center in the clumpy model \citep{Gronke16_model}.

\section{Discussion}\label{sec:discussion}

\subsection{Interpretation of Model Parameters}\label{sec:Interpretation}
Assuming that the degeneracy we present in \S\ref{sec:degeneracy} can be somehow broken (see possible examples in the following \S\ref{sec:break}), it is of great interest to decipher the crucial physical properties (kinematics, column density, etc) of the \lya\ scattering gaseous medium encoded in observed \lya\ spectra, which exist ubiquitously in the Universe and often exhibit a diversity of morphology, e.g. different numbers and shapes of peaks, peak flux ratios, and peak separations. In this section, we summarize our findings on how one should interpret the parameters of the shell or clumpy model derived from fitting observed \lya\ profiles. We will focus on the ``very clumpy" regime ($f_{\rm cl} \gg f_{\rm cl,\,crit}$) unless otherwise noted.

\begin{enumerate}

\item \textbf{${\rm H}\,{\textsc {i}}$ column density}: This parameter can be constrained by the peak separation and the extent of the wings of the \lya\ profile. The best-fit shell model gives the \HI\ column density of the shell $N_{\rm HI,\,{\rm shell}}$, whereas the best-fit clumpy model gives the \emph{total} \HI\ column density within the clumps\footnote{The total \HI\ column density in the ICM is usually negligible compared to that within the clumps, as the ICM is usually much hotter ($\gtrsim 10^6$\,K) and has a much lower \HI\ number density.}, given by $N_{\rm HI,\,{\rm total}} = \frac{4}{3}f_{\rm cl}N_{\rm HI,\,cl}$. As we have shown in previous sections, $N_{\rm HI,\,{\rm shell}} \simeq N_{\rm HI,\,{\rm total}}$ usually holds for the clumpy model and the best-fit shell model, suggesting that the \HI\ column density can be robustly determined from fitting.

However, this parameter should be treated with at least two caveats: (1) As we have shown in \S\ref{sec:degeneracy} and \ref{sec:random}, in the optically thick regime, the (total) \HI\ column density is degenerate with the shell effective temperature (or the random velocity of the clumps). Therefore, in order to get a well-constrained \HI\ column density by fitting observed \lya\ spectra, additional constraints are needed to break the degeneracy; (2) As both the shell model and the clumpy model assume an isotropic \HI\ gas distribution, whereas in actual astrophysical environments the gas distribution is more likely to be anisotropic, the derived $N_{\rm HI}$ value should be regarded only as an average value along the paths of escape of the \lya\ photons (which is actually not necessarily the average column density either along the line-of-sight or of all angles).

\item \textbf{Shell effective temperature (or Doppler parameter) / Clump velocity dispersion}: This parameter can be constrained by the width of the \lya\ profile, i.e. the FWHM of the peak(s). As we have shown, the shell effective temperature is usually equal to the effective temperature of the clumpy model with velocity dispersion $\sigma_{\rm cl}$ (see Eq. \ref{eq:Teff}). In other words, the turbulent velocity term in the Doppler parameter of the shell model (see Eq. 3 in \citealt{Verhamme06}) is equivalent to $\sigma_{\rm cl}$ of the clumpy model for the same \lya\ profile.

We hereby highlight a scenario where the fitted $T_{\rm eff}$ of the shell model cannot be interpreted literally. If a \lya\ spectrum is very broad and has very extended wings, it may require a high $T_{\rm eff} \gtrsim 10^6 \rm\,K$\footnote{The FWHM of a \lya\ profile is positively correlated with $N_{\rm HI,\,shell}$ and $T_{\rm eff}$; for a static sphere, FWHM $\simeq 320 \Big{(}\frac{N_{\rm HI,\,shell}}{10^{20}\,{\rm cm^{-2}}}\Big{)}^{1/3} \Big{(}\frac{T_{\rm eff}}{10^{4}\,{\rm K}}\Big{)}^{1/6}\,\rm km\,s^{-1}$ (see Eq. (87) from \citealt{Dijkstra17}).}, or equivalently, Doppler parameter $b \gtrsim 100\,\rm km\,s^{-1}$. Such a high Doppler parameter already corresponds to a very high internal turbulent Mach number $\mathcal{M}_{\rm cl}^{\rm turb} \sim 10$, which is enough to disintegrate the \HI\ shell or shock-ionize the \HI\ gas. Alternatively, we should interpret this as a clump velocity dispersion $\sigma_{\rm cl} \gtrsim 100\,\rm km\,s^{-1}$ of the clumpy model, which is physically reasonable in a strong gravitational field and/or in the presence of feedback. 

\item \textbf{Shell expansion / Clump outflow velocity}: This parameter can be constrained by the red-to-blue peak flux ratio of the \lya\ profile. As we have shown, $v_{\rm exp} \simeq v_{\rm cl}$ usually holds for the clumpy model and the best-fit shell model, suggesting that this parameter can also be robustly determined from fitting. However, the fitted shell expansion velocity ($v_{\rm exp}$) should not be interpreted literally as the bulk outflowing velocity of the \HI\ gas in at least two cases: (1) The actual velocity field of the \HI\ gas varies spatially. For example, in \S\ref{sec:outflow} we find that the best-fit shell model to a clumpy slab model with either a linearly increasing outflow or a ``momentum-driven + gravitational deceleration'' velocity profile has $v_{\rm exp} \simeq \overline{v_{\rm cl}(r)}$. Moreover, if the UV absorption lines suggest a series of outflow velocities, $v_{\rm exp}$ is expected to lie between the minimum and maximum absorption velocities; (2) The fitted \lya\ spectrum emerges from a multiphase scattering medium. For example, in \S\ref{sec:multiphase_slab} we show that the best-fit shell model to a multiphase clumpy slab model with cool clumps outflowing at $v_{\rm cl}$ and a hot ICM outflowing at $v_{\rm ICM}$ has $v_{\rm exp} < max\{v_{\rm cl}, v_{\rm ICM}\}$ unless $v_{\rm cl} = v_{\rm ICM}$. Therefore, the fitted $v_{\rm exp}$ should be interpreted as an average outflow velocity -- both space-wise and phase-wise.

\item \textbf{Intrinsic Line Width}: This parameter can be constrained by the extent of the wings of the \lya\ profile (for the shell model only; in the clumpy model, the intrinsic line width is fixed to be small and $\sigma_{\rm cl}$ is responsible for the broadening of the wings). It is well known that the fitted intrinsic line width $\sigma_{\rm i}$ of the shell model is usually overly large compared to the widths of the observed Balmer lines \citep{Orlitova18}. In this work, we have shown in \S\ref{sec:random} that a large intrinsic line width $\sigma_{\rm i}$ is always required for a shell model to fit a clumpy slab model with randomly moving clumps, and (1) $\sigma_{\rm i}$ increases as $\sigma_{\rm cl}$ increases; (2) $\sigma_{\rm i} > \sigma_{\rm cl}$ always holds (see Figure \ref{fig:sigma}). This is due to the intrinsic difference between a shell model and a clumpy model with corresponding parameters: compared to the shell model, the clumpy model naturally has more extended wings and lower but more extended peaks (see Figure \ref{fig:small_sigmai}), and is better suited for fitting broad \lya\ spectra with extended wings. This $\sigma_{\rm cl} < \sigma_{\rm i}$ trend, together with the quadruple degeneracy that we have discussed in \S\ref{sec:degeneracy}, provides a viable solution to the three major discrepancies emerged from shell model fitting as reported by \citet{Orlitova18}. It also suggests that the large $\sigma_{\rm i}$ values required in shell model fitting may simply imply a clumpy gas distribution (with a considerable velocity dispersion).

\item \textbf{Systemic Redshift}: When fitting an observed \lya\ profile, a parameter that dictates the systemic redshift of the modeled \lya\ source is usually introduced in post-processing. As \lya\ profile fitting is usually done in velocity space, this parameter can be specified as ${\Delta v}$, which is the difference between the systemic velocity of the modeled \lya\ source and the zero velocity of the observed \lya\ profile. For a typical double-peak \lya\ profile with a central trough between two peaks, the ${\Delta v}$ of the best-fit shell model is correlated with $v_{\rm exp}$, as the optical depth is maximum at $\sim -v_{\rm exp}$ (i.e. the trough location; see \citealt{Orlitova18}). In other words, ${\Delta v}$ and $v_{\rm exp}$ are intrinsically degenerate with each other.

Now the clumpy model offers us more possibilities to solve this issue with more flexibility. Although a single-phase slab with a very high clump covering factor ($f_{\rm cl} \gg f_{\rm cl,\,crit}$) basically converges to the shell model, a multiphase clumpy medium can produce many different trough shapes: for example, for $f_{\rm cl} \gg f_{\rm cl,\,crit}$ with a static or outflowing ICM, the trough can extend to both sides of the zero velocity (see Figure \ref{fig:vICM_fits}); for $f_{\rm cl} \simeq f_{\rm cl,\,crit}$ with a static ICM, the trough has residual flux and is always located at the line center. More modeling of observed \lya\ profiles is needed to examine whether these possibilities are physically reasonable.
\end{enumerate}

\subsection{Breaking the Degeneracy}\label{sec:break}
The intrinsic parameter degeneracy of the shell model (and the clumpy model as well, at least in the ``very clumpy'' regime) that we have described in \S\ref{sec:degeneracy} concerns us that how much meaningful physical information, if any, can be extracted from \lya\ spectra via RT modeling. In this section, we speculate several scenarios where the intrinsic parameter degeneracy can be broken and the physical properties of the \lya\ scattering medium can actually be constrained. 

\begin{enumerate}

\item \textbf{An accurate measurement of the systemic redshift of the \lya\ source}: Assuming that all the \lya\ photons are generated from recombination and nebular emission line(s) are clearly detected (e.g. \ha, \hb, or \oiii), the systemic redshift (i.e. the $\Delta v$ parameter) of the \lya\ source can be constrained reasonably well, and hence breaks the degeneracy. However, this requires that: (1) the observed \lya\ spectrum has a clear trough between the double peaks so that $v_{\rm exp}$ can be constrained; (2) the asymmetry of the observed \lya\ spectrum is significant enough so that the corresponding $v_{\rm exp}$ is much higher that the uncertainty of $\Delta v$. 

\item \textbf{Additional observational constraints on the gas outflow velocity / velocity dispersion / {\rm H\,{\textsc {i}}} column density}: If additional information is available from other observations, it may also help break the parameter degeneracy. Nevertheless, as such quantities are derived rather than directly observed (e.g. the gas outflow velocity can be deduced from UV absorption lines, and the gas velocity dispersion can be inferred from the widths of nebular emission lines), it is more reasonable to treat them as priors that confine the parameter space. Therefore, unless these additional constraints are reasonably stringent, the output parameters will still suffer from the degeneracy (which actually exists continuously across the parameter space).

\item \textbf{The Ly$\alpha$ profile corresponds to an optically thin regime}: As we have only found the parameter degeneracy in the optically thick regime, it is anticipated that if the Ly$\alpha$ profile does not belong to this regime, it may not be heavily affected by the parameter degeneracy. \lya\ spectra emerged from \HI\ with very low column densities ($\lesssim 10^{18}\,{\rm cm}^{-2}$) will be naturally in the optically thin regime -- they often exhibit narrow peak separations and/or residual flux at line center. However, objects that produce such \lya\ profiles are presumably LyC leakers and are rare in the Universe \citep{Cooke14, Verhamme15}.

In short, one should be cautious when interpreting the extracted parameters from fitting observed \lya\ spectra with idealized RT models. Additional observations on other lines may help break the intrinsic parameter degeneracy and better constrain the properties of the gaseous medium, although sometimes different types of constraints may contradict each other and yield unsuccessful fits (see e.g. Section 4.1 in \citealt{Orlitova18}). In that case, development of more advanced RT models that are more physically realistic and flexible may help solve this issue in the future.

\end{enumerate}

\section{Conclusions}\label{sec:conclusion}
In this work, we have explored what physical properties can be extracted from \lya\ spectra via radiative transfer modeling. The main conclusions of this work are:

\begin{enumerate}

\item Intrinsic parameter degeneracies exist in the widely-used shell model in the optically thick regime. For static shells, models with the same $N_{\rm HI,\,shell}\,T_{\rm shell}^{0.5}$ exhibit nearly identical \lya\ spectra. For outflowing shells, a quadruple degeneracy exists among $(v_{\rm exp}, N_{\rm HI,\,shell}, T_{\rm shell}, \Delta v)$. This finding reveals the limitations of the shell model and cautions against making any reasonable statements about the physical properties of the \lya\ scattering medium with only shell model fitting (cf. \S\ref{sec:degeneracy});

\item The parameters of a ``very clumpy'' slab model have a close correspondence to the parameters of the shell model. Specifically, (1) the \emph{total} column density of the clumpy slab model, $N_{\rm HI,\,{\rm total}} = \frac{4}{3}f_{\rm cl}N_{\rm HI,\,cl}$ is equal to the \HI\ column density of the shell model, $N_{\rm HI,\,shell}$; (2) the effective temperature of the clumpy slab model, $T_{\rm eff, slab} = T_{\rm cl} + \frac{\sigma_{\rm cl}^2 m_{\rm H}}{2 k_{\rm B}}$, where $\sigma_{\rm cl}$ is the 1D velocity dispersion of the clumps, is equal to the effective temperature of the shell model, $T_{\rm shell}$; (3) the average radial clump outflow velocity, $\overline{v_{\rm cl}(r)}$, is equal to the shell expansion velocity, $v_{\rm exp}$. This reminds us that the shell model parameters should be interpreted in a more physically realistic context rather than literally; 

\item In the shell model, large intrinsic line widths (several times of $\sigma_{\rm cl}$) are required to reproduce the wings of the clumpy slab models, reflecting the intrinsic difference between two different models. This $\sigma_{\rm cl} < \sigma_{\rm i}$ trend, together with the quadruple degeneracy, provides a viable solution to the three major discrepancies emerged from shell model fitting as reported by \citet{Orlitova18};

\item Adding another phase of hot inter-clump medium to the clumpy slab model will increase peak separation and boost $T_{\rm shell}$, but keeps $N_{\rm HI,\,shell} \simeq N_{\rm HI,\,total}$. The fitted $v_{\rm exp}$ lies between $v_{\rm cl}$ and $v_{\rm ICM}$. In particular, if $v_{\rm cl} = v_{\rm ICM}$, i.e., the cool clumps and the hot ICM are co-outflowing at the same speed, we get $v_{\rm exp} \simeq v_{\rm cl} = v_{\rm ICM}$;

\item For multiphase, clumpy spherical models, if $f_{\rm cl}$ is much larger than a critical value $f_{\rm cl,\,crit}$, the parameter correspondence still holds, albeit with larger uncertainties due to the geometrical difference; whereas if $f_{\rm cl} \simeq f_{\rm cl,\,crit}$, no direct correlation has been found between the shell and clumpy model parameters.

\end{enumerate}

In general, in order to obtain meaningful constraints on the physical properties of the \lya\ scattering gaseous medium, one should try to break the intrinsic parameter degeneracies revealed in this work with extra information from additional observations, rather than merely rely on fitting observed \lya\ spectra with idealized RT models. Moreover, the model parameters derived from \lya\ spectra fitting should not be understood literally -- instead, they should be interpreted in a more physically realistic context, e.g. in a multiphase, clumpy medium that we have explored in this work. Efforts in building more advanced RT models (e.g. with more realistic geometries) will also be helpful in the future.

\section*{Data availability}
The data underlying this article will be shared on reasonable request to the corresponding author. 

\section*{Acknowledgements}

We thank Phil Hopkins for providing computational resources. MG was supported by NASA through the NASA Hubble Fellowship grant HST-HF2-51409 awarded by the Space Telescope Science Institute, which is operated by the Association of Universities for Research in Astronomy, Inc., for NASA, under contract NAS5-26555. 
MG thanks the Max Planck Society for support through the Max Planck Research Group. Numerical calculations were run on the Caltech compute cluster ``Wheeler,'' allocations from XSEDE TG-AST130039 and PRAC NSF.1713353 supported by the NSF, and NASA HEC SMD-16-7592. We also acknowledge the use of the the following software packages: Astropy \citep{Astropy18}, the SciPy and NumPy system \citep{Scipy20, Numpy20}.

\bibliographystyle{mnras}
\bibliography{Clumpy}

\appendix

\section{Effect of ICM temperature on \lya\ model spectra}\label{sec:Ticm_spec}
Here we show that adding a hot phase of ICM does not necessarily affect the \lya\ model spectrum unless it satisfies a certain condition. Specifically, the transmission function $\mathcal{T}(x) = e^{-\tau(x)} = e^{-\sigma_{\rm HI}(x, T)\,N_{\rm HI}}$ needs to be wider for the hot phase than the cool phase.

In the core of the \lya\ line, the \HI\ cross section, $\sigma_{\rm HI} = \sigma_0 H(a_v, x) \sim\ \sigma_0 {e^{-x^2}}$, where $H(a_v, x)$ is the Voigt function, and $\sigma_0 \approx 5.895\times 10^{-14} (T / 10^4\,{\rm K})^{-1/2} \rm cm^{-2}$ is the \HI\ cross section at line center. Assuming that at a certain optical depth $\tau$, $\mathcal{T}(x)$ becomes sufficiently small and reaches a threshold $\mathcal{T}_{0}$ (e.g. for $\tau(x) \gtrsim 7$, $\mathcal{T}(x) \lesssim 10^{-3}$), and denoting $C_1$ = $5.895\times 10^{-12} \rm cm^{-2}$ so that $\sigma_0 = C_1/\sqrt{T}$, we have:

\begin{equation}
e^{-\frac{C_1}{\sqrt{T}}e^{-x^2}N_{\rm HI}} = \mathcal{T}_{0}
\label{eq:TICM1}
\end{equation}
so the threshold frequency can be solved as:
\begin{equation}
x_0 = \sqrt{-\ln\Bigg{(}-{\frac{\sqrt{T}}{C_1 N_{\rm HI}}\ln \mathcal{T}_{0}\Bigg{)}}}
\label{eq:TICM2}
\end{equation}
which corresponds to a threshold velocity:
\begin{equation}
v_0 = x_0 v_{\rm th}= \sqrt{-\ln\Bigg{(}-{\frac{\sqrt{T}}{C_1 N_{\rm HI}}\ln \mathcal{T}_{0}\Bigg{)}}} \sqrt{\frac{2k_{\rm B}T}{m_{\rm H}}}
\label{eq:TICM3}
\end{equation}

\begin{figure}
\centering
\includegraphics[width=0.5\textwidth]{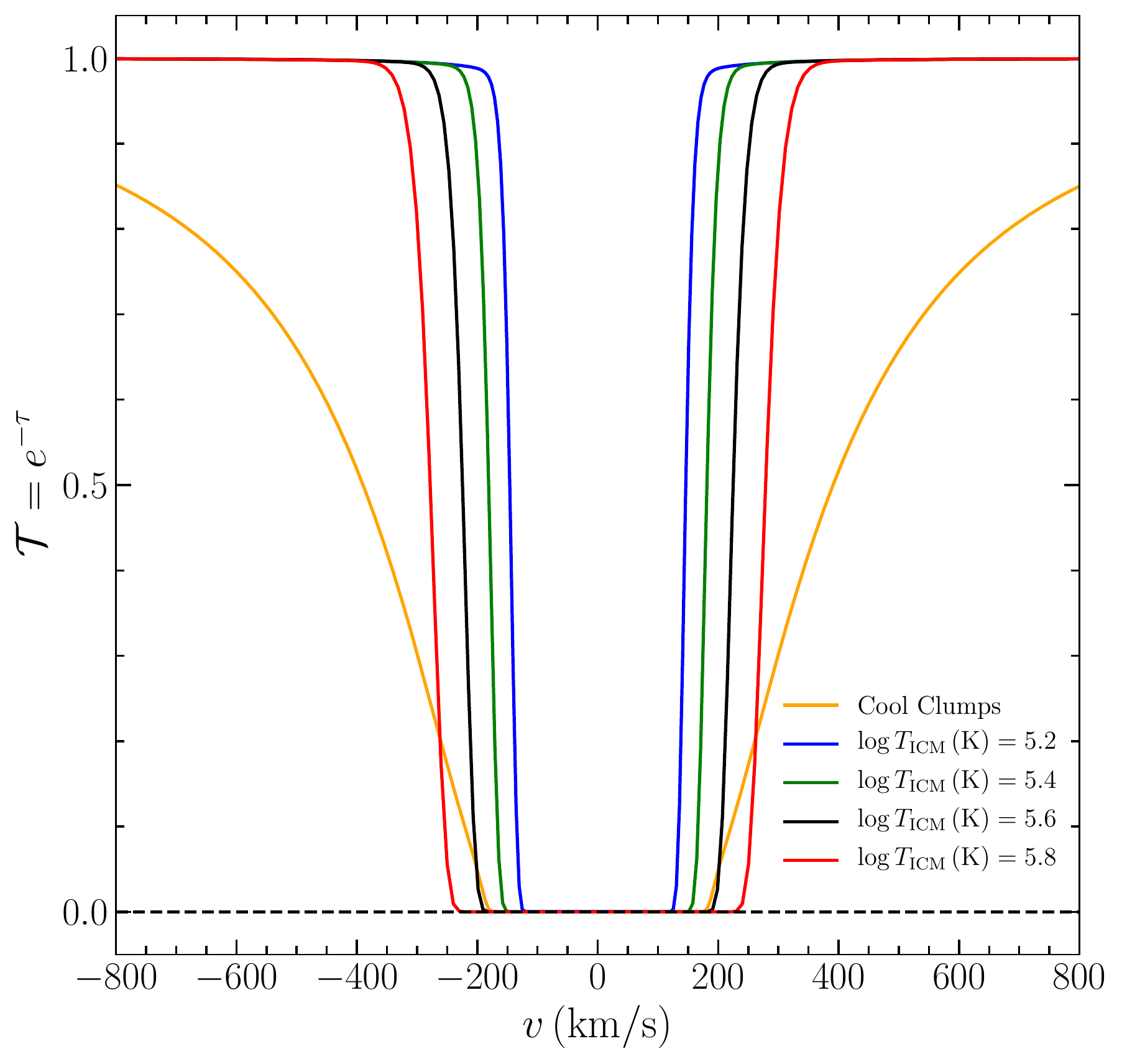}
\includegraphics[width=0.5\textwidth]{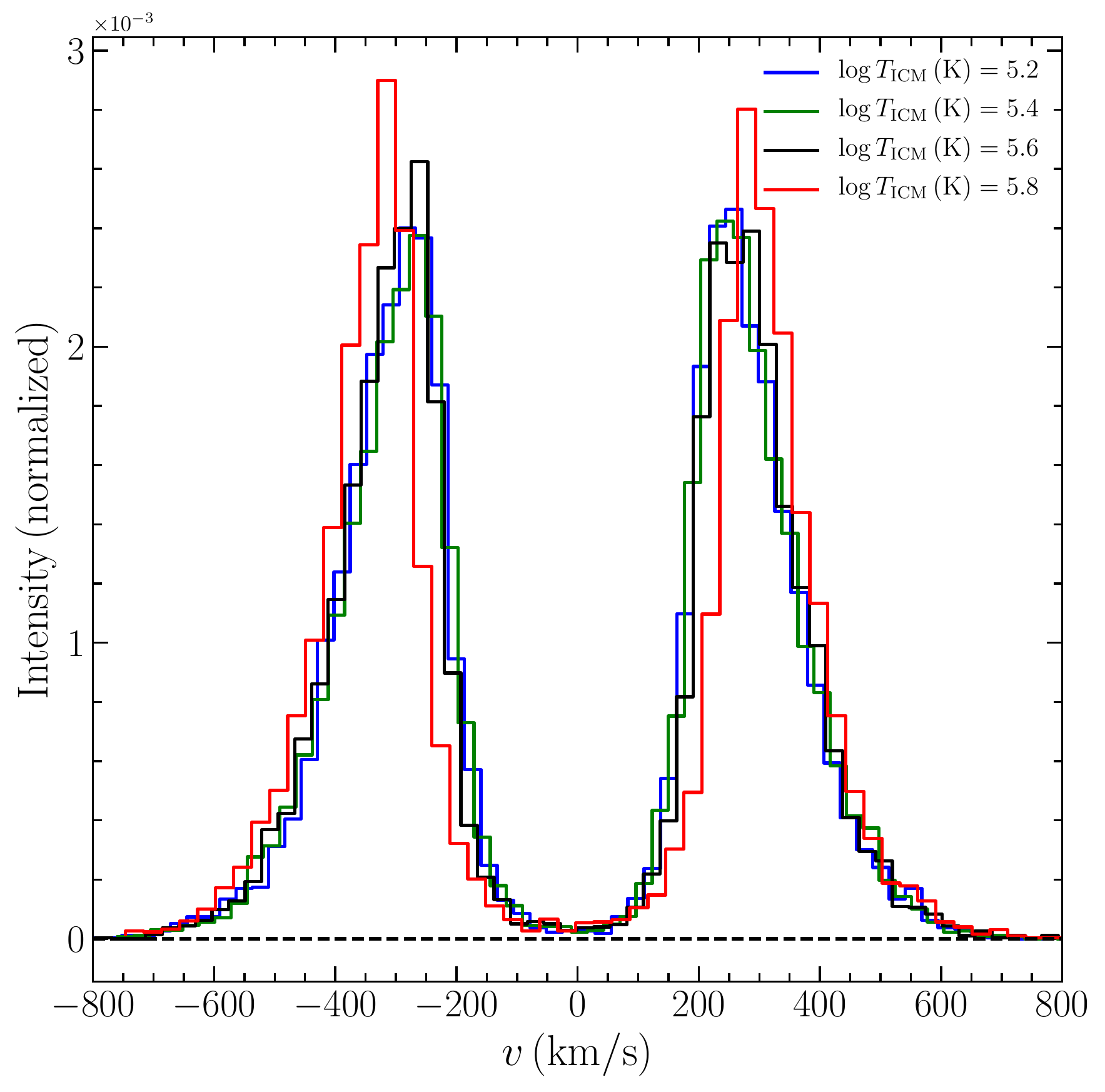}
    \caption{\textbf{Effect of ICM temperature on \lya\ transmission function and \lya\ model spectra.} Here we show one example: a two-phase scattering medium that consists of cool clumps with velocity dispersion $\sigma_{\rm {\rm cl}} = 50\,\rm km\,s^{-1}$ (hence effective temperature ${\rm log}\,T_{\rm eff}({\rm K}) = 5.2$) and total \HI\ column density ${\rm log}\,N_{\rm HI,\,total,\,cool} = 19.6$, and a hot ICM with total \HI\ column density ${\rm log}\,N_{\rm HI,\,total,\,ICM} = 17.2$. \emph{Top:} The transmission function of the cool clumps (the orange curve) as compared to those of the ICM at different temperatures. \emph{Bottom:} The model \lya\ profile as a function of the ICM temperature. As inferred from Eq. (\ref{eq:TICM3}), at ${\rm log}\,T_{\rm ICM}({\rm K}) \sim 5.6$ (the black curves) the ICM starts to have an impact on both the transmission function and the model \lya\ profile.
    \label{fig:TICM}}
\end{figure}

In order to have an impact on the \lya\ profile, the hot phase of ICM needs to have a threshold velocity larger than that of the cool phase, i.e. $v_{\rm 0,\,ICM} > v_{\rm 0,\,cool}$. We hereby consider the following example: a two-phase scattering medium that consists of cool clumps with velocity dispersion $\sigma_{\rm {\rm cl}} = 50\,\rm km\,s^{-1}$ (hence effective temperature ${\rm log}\,T_{\rm eff}({\rm K}) = 5.2$) and total \HI\ column density ${\rm log}\,N_{\rm HI,\,total,\,cool} = 19.6$, and a hot ICM with total \HI\ column density ${\rm log}\,N_{\rm HI,\,total,\,ICM} = 17.2$. Using Eq. (\ref{eq:TICM3}) and demanding $v_{\rm 0,\,ICM} > v_{\rm 0,\,cool}$ yields that ${\rm log}\,T_{\rm ICM}({\rm K}) \gtrsim 5.6$ is required for the ICM to have a wider transmission function than the cool clumps, and hence have a visible impact on the \lya\ profile. We illustrate this result in Figure \ref{fig:TICM} by showing a series of ICM transmission functions and model \lya\ profiles with different $T_{\rm ICM}$ values. It can be clearly seen that at ${\rm log}\,T_{\rm ICM}({\rm K}) \sim 5.6$ the ICM starts to have an impact on both the transmission function and the model \lya\ profile.
 
\section{Effect of ICM on the critical covering factor}\label{sec:Ticm_fccrit}
In \citet{Gronke17b}, an important physical quantity is defined -- the critical covering factor of the clumps, $f_{\rm cl,\,crit}$. It is the critical average number of clumps per line-of-sight, above which the clumpy scattering medium will behave like a homogeneous medium (i.e. a homogeneously filled shell or slab), and below which a significant number of \lya\ photons will escape near the line center. In this section, we test how much impact the hot ICM component has on $f_{\rm cl,\,crit}$, and hence on the boundaries of different RT regimes.

The value of $f_{\rm cl,\,crit}$ sets the transition between two physical regimes of \lya\ resonant scattering. Assuming that the ensemble of the clumps is optically thick at the \lya\ line center (which is always true throughout this work), if $f_{\rm cl} \lesssim f_{\rm cl,\,crit}$, photons scatter off the clumps in a random-walk manner, and the number of clumps a photon intercepts scales as $N_{\rm cl} \propto f_{\rm cl}^2$; whereas if $f_{\rm cl} \gtrsim f_{\rm cl,\,crit}$, photons escape via a frequency excursion (i.e. a series of wing scatterings), and the number of clumps a photon intercepts scales as $N_{\rm cl} \propto f_{\rm cl}$. Therefore, $f_{\rm cl,\,crit}$ can be estimated by determining the turning point of the scaling relation between $N_{\rm cl}$ and $f_{\rm cl}$ (see Figures 2, 4 and 6 from \citealt{Gronke17b}).

\begin{figure}
\centering
\includegraphics[width=0.4\textwidth]{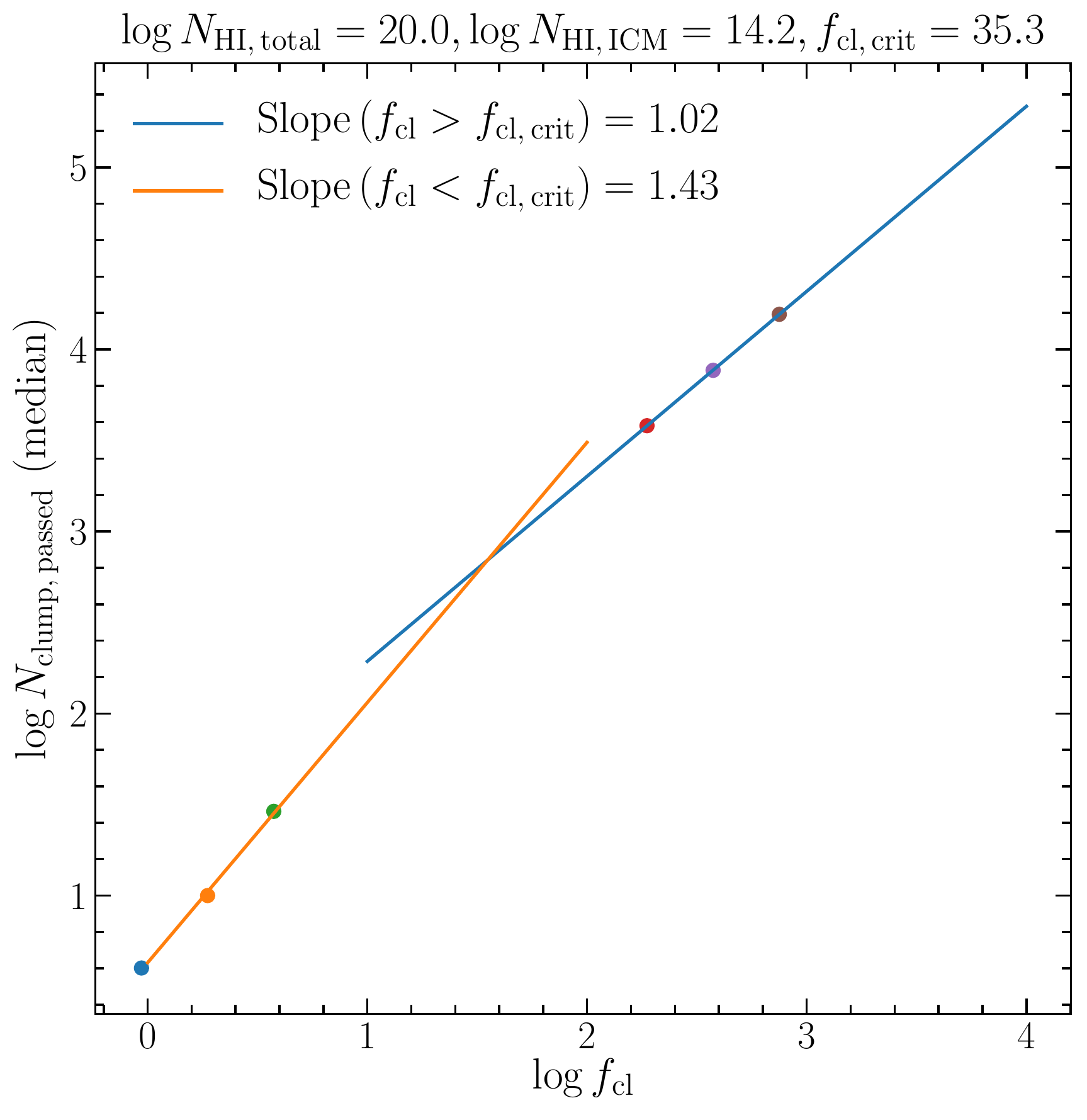}
\includegraphics[width=0.4\textwidth]{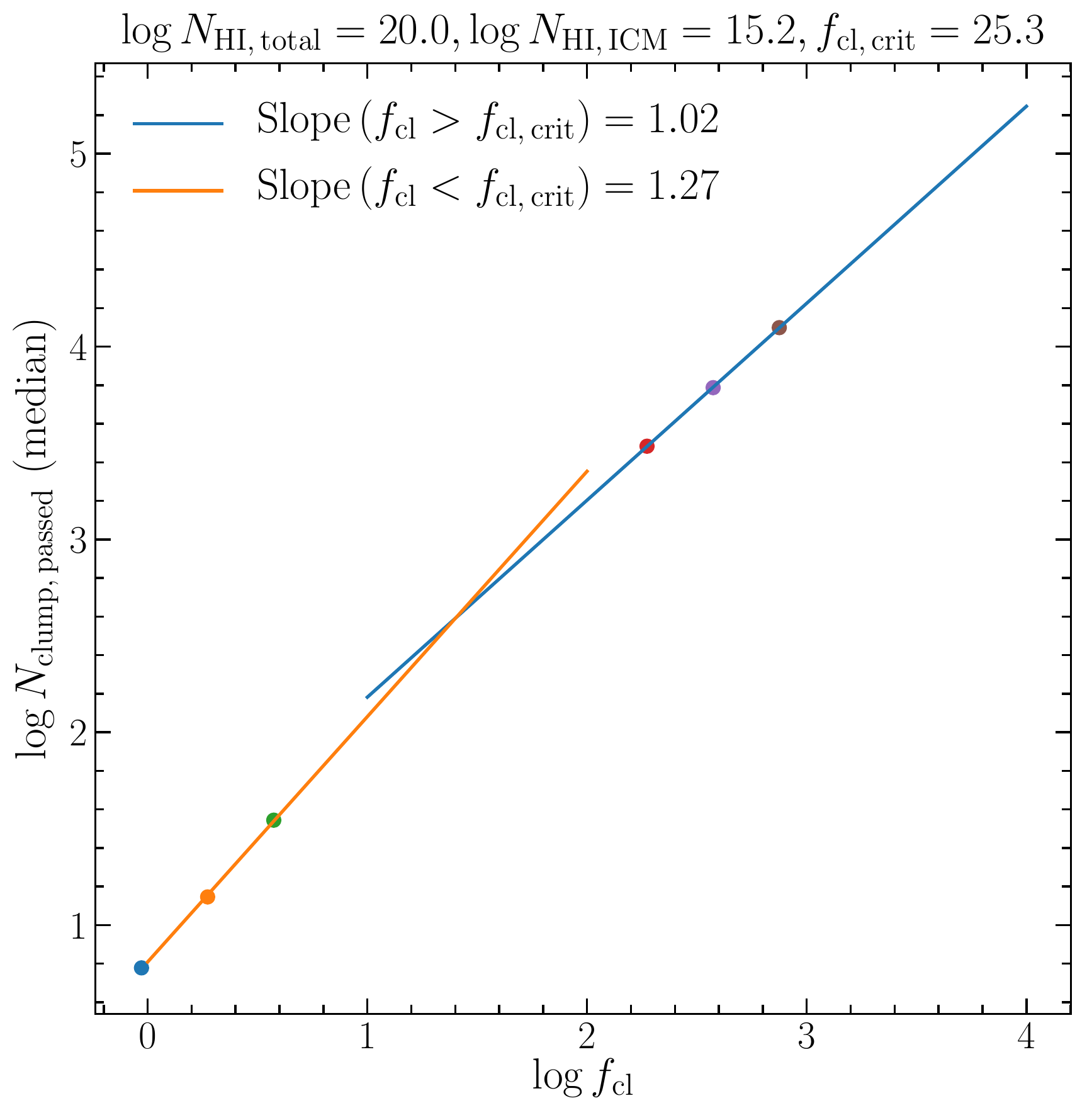}
\includegraphics[width=0.4\textwidth]{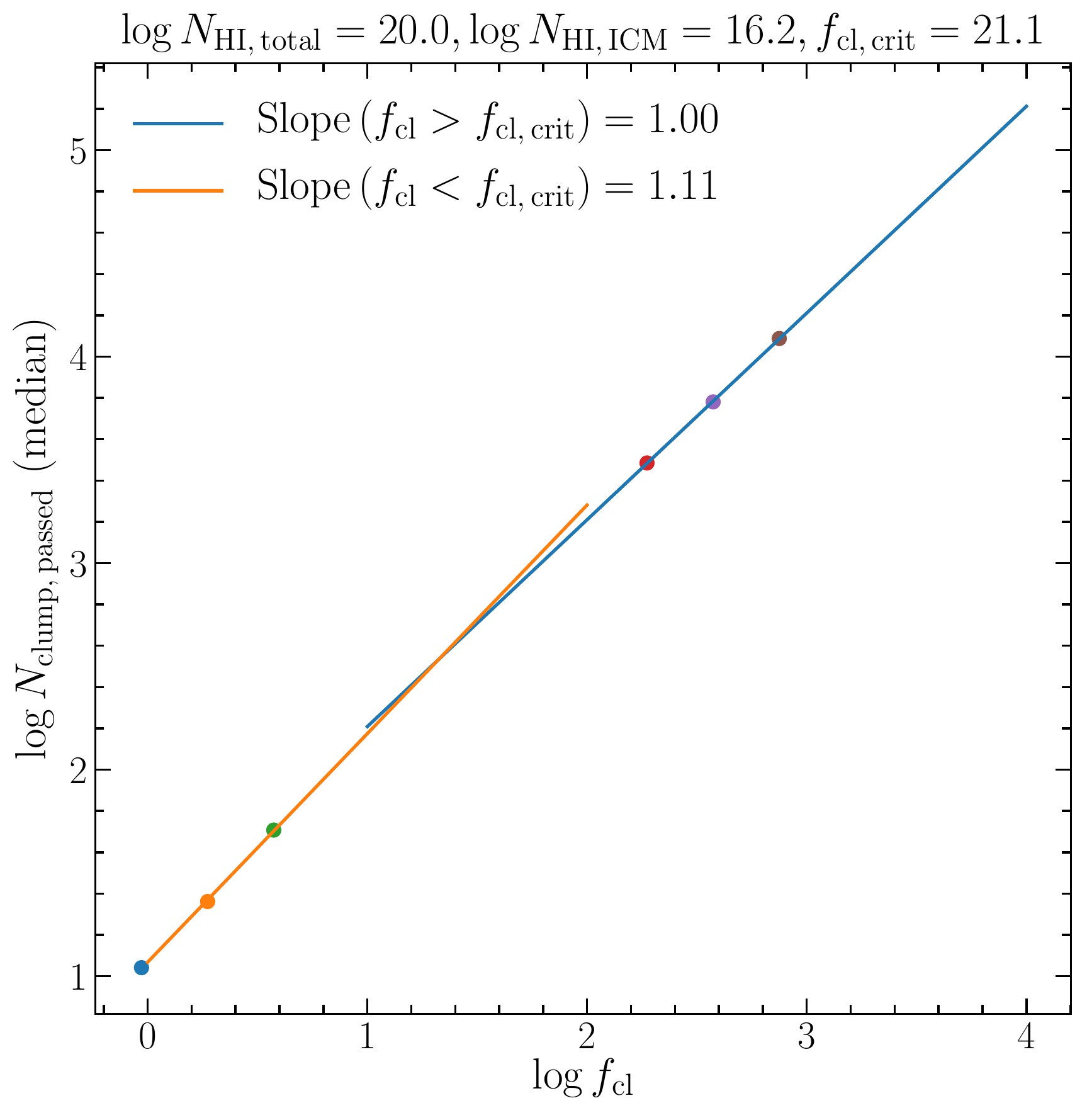}
    \caption{\textbf{Effect of ICM with different column densities on the critical clump covering factor, $f_{\rm cl,\,crit}$.} Here we show one set of examples: a two-phase clumpy slab with ${\rm log}\,N_{\rm HI,\,total} = 20.0$ in the static clumps, and a hot ICM with total \HI\ column density ${\rm log}\,N_{\rm HI,\,ICM} = 14.2 - 16.2$ (or equivalently, $n_{\rm HI,\,{\rm ICM}} = 10^{-6} - 10^{-4}\,\rm cm^{-3}$). With ${\rm log}\,N_{\rm HI,\,ICM}$ varying by two orders of magnitudes, $f_{\rm cl,\,crit}$ only changes by a factor of $\sim$ 1.5, suggesting that the hot ICM only has a minor effect on $f_{\rm cl,\,crit}$ and the boundaries of different RT regimes.
    \label{fig:fccrit_ICM}}
\end{figure}

Here we show one set of examples in Figure \ref{fig:fccrit_ICM}: a two-phase clumpy slab with ${\rm log}\,N_{\rm HI,\,total} = 20.0$ in the clumps (which are static), and a hot ICM with total \HI\ column density ${\rm log}\,N_{\rm HI,\,ICM} = 14.2 - 16.2$ (or equivalently, $n_{\rm HI,\,{\rm ICM}} = 10^{-6} - 10^{-4}\,\rm cm^{-3}$). It can be seen that with ${\rm log}\,N_{\rm HI,\,ICM}$ varying by two orders of magnitudes, $f_{\rm cl,\,crit}$ only changes by a factor of $\sim$ 1.5. This result suggests that although under certain conditions, the hot ICM can have a significant impact on the model \lya\ spectrum (see \S\ref{sec:multiphase_slab} and \S\ref{sec:multiphase_sphere}), it only has a minor effect on $f_{\rm cl,\,crit}$ and the boundaries of different RT regimes.

\bsp	
\label{lastpage}
\end{document}